\documentclass[prd,nofootinbib,notitlepage]{revtex4-1}
\usepackage{epsfig,amsmath,slashed,subfigure}

\newcommand{\bra}[1]{\langle #1|}
\newcommand{\ket}[1]{|#1\rangle}
\newcommand{\braket}[2]{\langle #1|#2\rangle}

\def\Nj{\{N_j\}}

\def\d{{\rm d}}
\def\e{{\rm e}}
\def\i{i}
\def\parti{n} 
\def\Pro{{\sf P}}

\newcommand{\tr}{{\rm tr}}  
  
\newcommand{\uq}{{\rm u}}
\newcommand{\dq}{{\rm d}}
\newcommand{\sq}{{\rm s}}

\newcommand{\Qf}{\mathbf{Q}}

\newcommand{\sigmav}{\boldsymbol{\sigma}}
\newcommand{\iotav}{\boldsymbol{\iota}}
\newcommand{\rhov}{\boldsymbol{\rho}}

\newcommand{\muv}{\boldsymbol{\mu}}
\newcommand{\gs}{\gamma_S}
\newcommand{\p}{{\rm p}}

\newcommand{\ee}{$\textrm{e}^+\textrm{e}^-$~}
\newcommand{\ppb}{$\textrm{p}\overline{\textrm{p}}$}

\def\NP{{ Nucl.\ Phys.\ }}
\def\PL{{ Phys.\ Lett.\ }}

\begin{document}

\title{Statistical hadronization with exclusive channels in \ee annihilation}

\author{L. Ferroni}
\affiliation{Institut f\"ur Theoretische Physik, Johann Wolfgang 
Goethe-Universit\"at, Frankfurt am Main, Germany}
\author{F. Becattini}
\affiliation{Universit\`a di Firenze and INFN Sezione di Firenze, Florence,
Italy}

\begin{abstract}
We perform a systematic analysis of exclusive hadronic channels in \ee 
collisions at centre-of-mass energies between 2.1 and 2.6 GeV within the 
statistical hadronization model. Because of the low multiplicities involved, 
calculations have been carried out in the full microcanonical ensemble, 
including conservation of energy-momentum, angular momentum, parity, 
isospin, and all relevant charges. We show that the data is in an overall good 
agreement with the model for an energy density of about 0.5 GeV/fm$^3$ 
and an extra strangeness suppression parameter $\gamma_S \sim 0.7$, 
essentially the same values found with fits to inclusive multiplicities at 
higher energy.
\end{abstract}

\maketitle

\section{Introduction}

The statistical approach to multi-hadron production \ee annihilations has a long story.
Early works date back to the '70s \cite{eestat,eestat2}, with different versions 
of the model and different observables examined in the relevant analyses, such as inclusive
yields, multiplicity distributions etc. All of these calculations involved simplifying
assumptions and drastic approximations, mostly because of the lack of computing power, 
so that in practice it was very difficult to confirm or rule out {\em the} statistical 
model, also in view of it being conceived as a full alternative to a dynamical model.
Nowadays, with QCD being the accepted theory of strong interactions, the Statistical 
Hadronization Model (SHM) has resurged as a model of hadronization, it has a framework 
based on quantum statistical mechanics \cite{review}, and it has been extensively and 
succesfully applied to the analysis of inclusive multiplicities in elementary \cite{vari} 
and relativistic heavy ion collisions \cite{varih}. Also, the SHM was shown 
to succesfully reproduce transverse momentum spectra in hadronic collisions \cite{becagp} 
with specific predictions concerning the approximately exponential shape of the 
low-$p_T$ spectrum and the so-called $m_T$ scaling phenomenon. 
This model has by now become a standard tool in heavy ion physics while the reasons of 
its success in elementary collisions are still subject of debate \cite{diba}. 

It is a common belief that statistical equilibrium cannot be attained in elementary
collisions via post-hadronization collisions because of the low multiplicities
and the rapid expansion. Therefore, one is led to conclude that statistical equilibrium 
is an intrinsic feature of the hadronization process itself, as envisaged by Hagedorn 
many years ago ("hadrons are born at equilibrium" \cite{hage}). In the latter case, 
two possibilities arise:
\begin{itemize}
\item{-} the apparent statistical equilibrium is just mimicked by a special 
property of the dynamics governing the hadronization tending to evenly populates 
all final states, but which has essentially nothing to do with a proper statistical 
system, which can be realized only within a finite volume.
\item{-} the apparent statistical equilibrium is established within a finite 
volume and therefore is a ''genuine" one; if the volume was large enough, a proper 
temperature could be introduced.
\end{itemize}

The former picture can be defined as {\em phase space dominance} to discriminate
it from proper statistical equilibrium. While phase space dominance is, though,
a highly non-trivial hypothesis, its predictions do quantitatively differ from the 
proper SHM. As pointed out in refs. \cite{meaning,review}, the dynamical
matrix element of the decay of a massive cluster into $N$ particles contain, in
the SHM case, peculiar quantum statistics terms (Bose-Einstein and Fermi-Dirac 
correlations) owing to the finite cluster volume, which are generally absent in the 
phase space dominance picture. The very fact that Bose-Einstein and Fermi-Dirac
correlations have been observed in elementary collisions demonstrates the finite 
extension of the hadron emitting source and therefore favours a model like the SHM
where finite volume is a built-in feature.

Anyhow, it would be desirable to quantitatively test the the genuine statistical
hadronization model on observables which are more sensitive to the form of the
matrix element than average multiplicities. For this purpose, in this work we 
compare the production rates of exclusive channels in \ee collisions at low energy 
with the predictions of the SHM. 
   
Exclusive channels in \ee collisions have been measured at low centre-of-mass energy 
($\lesssim 4$ GeV). At such a low energy, QCD is in the full non-perturbative regime
and one can assume that, unlike at higher energy where clusters are two or more 
(jettiness), only one hadronizing massive cluster at rest in centre-of-mass frame 
of the collision is forme. The price to be paid is that, in calculating the model 
predictions, none of the relevant conservation laws, including energy-momentum, 
intrinsic angular-momentum and parity, as well as internal symmetries, can be 
neglected (see e.g. ref.~\cite{heinzppb} where \ppb\, annihilation at rest was 
studied in the SHM). In the SHM framework, this means that one has to calculate 
averages in the most general microcanonical ensemble of the hadron-resonance gas. 

To carry out this calculation, in this work we take advantage of the formalism developed 
in two previous papers of ours \cite{bf3,bf4} where the microcanonical partition
function of an ideal multi-species relativistic gas was calculated enforcing the 
conservation of the maximal set of observables pertaining to space-time symmetries 
(energy-momentum, spin, helicity, parity). We extend the formulae obtained therein 
to the hadron-resonance gas including internal quantities conserved by strong interaction 
(isospin, C-parity and abelian charges). We then take into account resonance decays 
and compare the results of our calculations with the data collected in \ee collisions 
at low energy.

The paper is organized as follows: in Sect.~\ref{model} we will expound a formulation
of the SHM in the full microcanonical ensemble which is suitable for the problem
of exclusive channels and in Sect.~\ref{microchannel} we will obtain an expression
for the their rates; Sect.~\ref{numerics} will be focussed on a method to compute 
them numerically. Finally, Sect.~\ref{ee} will describe the analysis of data in \ee 
collisions at low energy will and Sect.~\ref{concl} will be devoted to a discussion
of the results and to conclusions.

\section{The Statistical Hadronization Model in the microcanonical ensemble}
\label{model}

\begin{figure}[htpb]
\begin{center}
\includegraphics[width=0.6\textwidth]{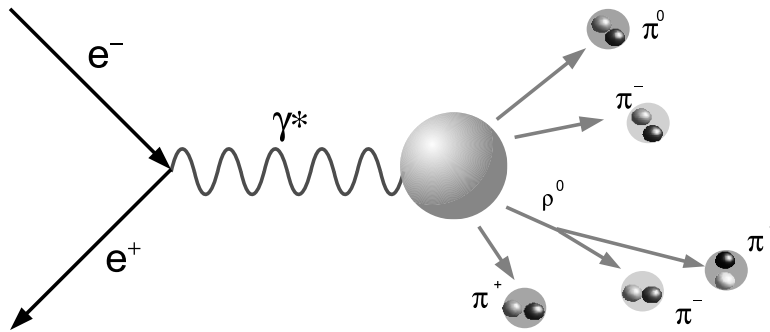}
\caption{\small{Dominant contribution to hadroproduction in a low energy \ee 
collision with formation of a single hadronizing cluster.}}
\label{eecoll}
\end{center}
\end{figure}

In the modern formulation of the SHM~\cite{review}, the strong interaction process
in a collision between particles leads to the formation of a set of extended massive 
objects called {\em clusters} or {\em fireballs}). Each cluster decays into hadrons 
in a purely statistical fashion, that is {\em any multi-hadronic state within the 
cluster compatible with its quantum numbers is equally likely}.  
The number of clusters produced, as well as their kinematical and internal quantum 
properties, are determined by the prior dynamical process and are not predictable 
within the SHM itself. Particularly, in high energy collisions ($\sqrt{s} \gtrsim 10$ 
GeV), the production of clusters following the perturbative parton shower stage leads 
to a multiple cluster production. Conversely, for energies sufficiently below the 
perturbative regime ($\sqrt{s} \lesssim 4$ GeV), one may expect that, to a very good 
approximation, a single cluster is formed (see fig.~\ref{eecoll}). Under this circumstance 
the whole centre-of-mass energy is spent to produce particles and no jet is observed. 
The cluster mass will then coincide with $\sqrt{s}$ and its relevant quantum 
numbers will be the same as those pertaining to the initial state of the collision.

Once the quantum numbers of a cluster are known, the assumption of equal probability
allows to perform calculations within the framework of statistical mechanics, in the
relativistic microcanonical ensemble. The use of the microcanonical ensemble is 
necessary as at such a low energy the effect of exact conservation laws is very 
important\cite{bf2}. It should be emphasized that the basic 
assumption of the model states that multi-particle {\em localized states} compatible 
with cluster's conserved quantities are equally likely, but these states do not 
coincide with observable free particle asymptotic states. Such a distinction is, 
for practical purposes, not an issue when the volume is sufficiently large, but it 
is relevant in principle and may result in quantitative differences when the volume 
is small, i.e. comparable with the third power of the typical lenght scale of the 
hadron world, the pion Compton wavelength ${\mathcal O}(1)$ fm$^3$. This is 
discussed in detail in refs.~\cite{meaning,review}.

If the cluster can be described as a mixture of states, the basic postulate implies 
that the corresponding density operator is a sum over all localized states projected 
onto the initial cluster's quantum numbers through a projection operator $\Pro_i$:
\begin{equation}\label{densmat}
\widehat \rho \propto \sum_{h_V} \Pro_i \ket{h_V} \bra{h_V} \Pro_i 
\equiv \Pro_i \Pro_V \Pro_i
\end{equation}
where $\ket{h_V}$ are multi-hadronic localized states and $\Pro_i$ is the projector 
onto the cluster's initial conserved quantities: energy-momentum, intrinsic angular 
momentum and its third component, parity and the generators of inner symmetries
of strong interactions \footnote{Operators in the Hilbert space will be denoted with
a hat. Exceptions to this rule are projectors, which will be written in serif font,
i.e. $\Pro$.}. If, on the other hand, the cluster is prepared in a pure quantum state 
$\ket{\psi}$(what is the case for a single produced cluster in \ee collisions) then, 
according to the basic assumption, this is ought to be an even superposition of all 
multi-particle localized states with the initial conserved quantities, that is:
\begin{equation}\label{pure}  
\ket{\psi} = \sum_{h_V} c_{h_V} \Pro_i \ket{h_V}  \qquad {\rm with}\,\, |c_{h_V}|^2=
 {\rm const}  \, .
\end{equation}
It can be readily shown \cite{review} that if the coefficients $c_{h_V}$ have random 
phases, an effective mixture description with the operator in (\ref{densmat}) is
recovered. Hence, a new hypothesis is introduced: if the cluster is a pure state, 
the superposition of multi-hadronic localized states must have random phases.

The operator $\Pro_i$ can be formally defined as the pseudo-projector (it is not
idempotent, see below) onto an irreducible vector of the full symmetry group and 
worked out in a group theory framework \cite{bf1,meaning,bf4}. It can be factorized 
into a "kinematic" pseudo-projector, associated to general space-time symmetries, 
and an actual idempotent projector for inner symmetries, associated to compact groups. 
For the general space-time symmetry the relevant group is the extended orthochronous 
Poincar\'e group IO(1,3)$^\uparrow$ and an irreducible state with positive mass is 
identified by a four-momentum $P$, a spin $J$ and its third component $\lambda$ and 
a discrete parity quantum number $\pi = \pm 1$. Therefore:
\begin{equation}\label{factor1}
   \Pro_i = \Pro_{P,J,\lambda,\pi} \Pro_{inner}  
\end{equation}
If the pseudo-projector $\Pro_{P,J,\lambda}$ is worked out in the cluster's rest frame
where $P=(M,{\bf 0})$, it further factorizes \cite{bf1,bf4}, i.e.:
\begin{equation}\label{poincare} 
 \Pro_{P,J,\lambda,\pi} = \delta^4(P - \hat P) \Pro_{J,\lambda} \Pro_\pi
 = \delta^4(P - \hat P) \Pro_{J,\lambda} \frac{{\sf I} + \pi \hat \Pi }{2} 
\end{equation}
where $\widehat P$ is the four-momentum operator, $\Pro_{J,\lambda}$ is a projector 
onto SU(2) irreducible states $\ket{J,\lambda}$ and $\hat{\Pi}$ is the space 
reflection operator. Thus, the pseudo-projector (\ref{factor1}) becomes:
\begin{equation}\label{factor2}
   \Pro_i = \delta^4(M - \widehat P^0) \delta^3(\widehat {\bf P})
   \Pro_{J,\lambda} \Pro_\pi \Pro_{inner}  
\end{equation}
Note that $\Pro_{J,\lambda}$, $\Pro_\pi$ and $\Pro_{inner}$ commute with each
other. 

As clusters are colour singlets by definition, the projector $\Pro_{inner}$
involves flavour and baryon number conservation. In principle, the largest symmetry 
group one should consider is SU(3) flavour, plus three other U(1) groups for 
baryon number, charm and beauty conservation. However, SU(3) symmetry is badly broken 
by the mass difference between strange and up, down quarks, so it is customary to 
take a reduced SU(2)$\otimes$U(1) where SU(2) is associated with isospin and U(1) 
with strangeness. The isospin SU(2) symmetry is explicitly broken as well, but the 
breaking term is small and can generally be neglected. However, most calculations 
in the past have replaced isospin SU(2) with another U(1) group for electric charge, 
so that the symmetry scheme, from an original 
SU(2)$_{\rm isospin} \otimes$U(1)$_{\rm strangeness}\otimes$U(1)$_{\rm baryon}$ 
reduces to U(1)$_{\rm charge}\otimes$U(1)$_{\rm strangeness} \otimes$U(1)$_{\rm baryon}$.

Altogether, $\Pro_{inner}$ can be written as
\begin{equation}
 \Pro_{inner} = {\Pro}_{I,I_3} {\Pro}_\Qf {\Pro}_\chi 
\end{equation}
where $I$ and $I_3$ are isospin and its third component, $\Qf = (Q_1,\ldots,Q_M)$ 
is a vector of $M$ integer abelian charges (baryon number, strangeness, etc.)
and $\Pro_\chi$ is the projector onto C-parity, which makes sense only if the system 
is completely neutral, i.e. $I=0$ and $\Qf = {\bf 0}$; in this case, ${\sf P}_\chi$ 
commutes with all other projectors. 
 
From the density operator (\ref{densmat}) the probability of observing an asymptotic 
multiparticle state $\ket{f}$ ensues:
\begin{equation}\label{prob}
 p_f = \frac{\bra{f} \Pro_i \Pro_V \Pro_i \ket{f}}{\tr (\Pro_i \Pro_V \Pro_i)}
\end{equation}
which is a well-defined one with regard to positivity and conservation laws because 
$p_f = 0$ if the state $\ket{f}$ 
has not the same quantum numbers as the initial state. The normalizing factor, i.e.
the trace of the operator $\Pro_i \Pro_V \Pro_i$ can be worked out as:
\begin{equation}\label{sumprob}
 \tr (\Pro_i \Pro_V \Pro_i) = \tr (\Pro^2_i \Pro_V) = \delta^4(0) 
 \tr (\Pro_i \Pro_V) \, .
\end{equation}
where we have used the particular form of $\Pro_i$ in eq.~(\ref{factor2}) and
we have taken into account that all operators except $\delta^4(P - \hat P)$ are 
idempotent.
The reason for the presence of a divergent positive constant $\delta^4(0)$ is the 
non-compactness of the Poincar\'e group, which makes in fact impossible to have a 
properly normalized projector. The last trace in (\ref{sumprob}) can be written as
\begin{equation}
 \tr (\Pro_i \Pro_V) = \sum_{h_V} \bra{h_V} \Pro_i \ket{h_V} \equiv \Omega
\end{equation}
which is, by definition the {\em microcanonical partition function} \cite{bf3}, 
i.e. the sum over all localized states projected onto the conserved quantities
defined by the selected initial state. If only energy and momentum conservation is 
enforced, $\Omega$ takes on a more familiar form:
\begin{equation}\label{mpf}
 \Omega = \sum_{h_V} \bra{h_V} \delta^4(P-\hat P) \ket{h_V}   \, .
\end{equation}

In principle, the asymptotic multi-particle states $\ket{f}$ in eq.~(\ref{prob}) only 
include strongly stable hadrons, while the interaction between them is understood
in the same equation through the projector $\Pro_i$ which contains the full hamiltonian
$\widehat P^0$ (see eq.~(\ref{factor2}). An outstanding theorem by Dashen, Ma and
Bernstein \cite{dmb} asserts that, in the thermodynamic limit, the partition function 
 - in any ensemble - of an interacting system is the sum of the partition 
function of the system without interaction and a term involving the scattering matrix 
between the otherwise free particles. The well-known consequence of this theorem 
is the so-called {\em hadron-resonance} gas model; if only the resonant part of the 
scattering matrix is retained (the background interaction is neglected), the 
interaction term of the partition function reduces to that of a gas of resonances
treated as free particles with distributed mass. 
Strictly speaking, there is an additional contribution from resonance interference, 
which might be sizeable in case of wide, overlapping resonances with the same decay 
channel, but this depends on mostly unknown complex parameters and is thus assumed
to vanish altogether or it is simply disregarded.

In the spirit of the DMB theorem and the hadron-resonance gas model, we will 
therefore calculate the probabilities (\ref{prob}) including resonances as free
particles with distributed mass in the multi-particle free states $\ket{f}$ and
let them decay afterwards. It must be stressed that this is an assumption going
beyond the scope of validity of the DMB theorem, which affirms the equality of two 
traces, and not of single trace terms. In other words, the hadron-resonance gas 
decomposition, strictly speaking, applies only to fully inclusive quantities and
not to partly inclusive like multiplicities of single species or exclusive
final states. Furthermore, the DMB theorem requires the thermodynamic limit $V\to
\infty$. Up to now, these problems have been ignored and the hadron-resonance 
gas model has been used to calculate hadron abundances and spectra when applying
the SHM to the data. As has been mentioned, we will continue to use the hadron-resonance 
gas model in its simplest form also for exclusive channel rates and for small 
clusters. This is likely to be a good approximation but it should be kept in mind 
that deviations may well be implied.

\section{The microcanonical channel weight}
\label{microchannel}

Denoting a multi-particle final state $\ket{f}$ as $\ket{ \Nj,\{ p \} }$ where 
$\Nj= (N_1, \ldots ,N_K)$ is the set of multiplicities $N_j$ for each particle
species $j$, i.e. the {\em channel}, and $\{ p \}$ stands for the set of kinematic 
variables (momenta and spin components or helicities) of the particles, we can 
calculate the probability of the channel $\Nj$ as:
\begin{eqnarray}\label{pchann}
 p_{\Nj} &=& \frac{1}{\delta^4(0) \Omega} 
 \sum_{\{ p \}} \bra{ \Nj,\{ p \}} \Pro_i \Pro_V \Pro_i \ket{\Nj,\{ p \}} \nonumber \\
  &=& \frac{1}{\delta^4(0) \Omega} \delta^4(0) 
  \sum_{\{ p \}} \bra{ \Nj,\{ p \}} \Pro_{J,\lambda} \Pro_\pi \Pro_{inner} \Pro_V 
  \Pro_i \ket{\Nj,\{ p \}}
\end{eqnarray}
where use has been made of the decomposition in eq.~(\ref{factor2}) and the fact 
that the final states are eigenstates of total energy-momentum. 
One of the difficulties of working out this expression in a relativistic quantum
field framework is that the localized states, in general, are {\em not} states with
a definite number of asymptotic particles, unlike in non-relativistic quantum mechanics
(see discussion in ref.~\cite{bf3}). However, since we sum over all kinematic states 
and the projectors $\Pro_{J,\lambda}$, $\Pro_\pi$ do not change the number of particles, 
i.e. the set $\Nj$, and we can use the ciclicity for these two projectors and rewrite 
the last expression as:
\begin{equation}\label{pchann2}
 p_{\Nj} = \frac{1}{\Omega} \sum_{\{ p \}} \bra{ \Nj,\{ p \}} \Pro_{inner} 
 \Pro_V \Pro_i  \Pro_\pi \Pro_{J,\lambda}\ket{\Nj,\{ p \}} 
\end{equation}
Now $\Pro_{inner}$ commutes with $\Pro_V$ because localization does not affect 
internal symmetries, as well as with $\Pro_i$, so we get:
\begin{eqnarray}\label{pchann3}
 p_{\Nj} = \frac{1}{\Omega} \sum_{\{ p \}} \bra{ \Nj,\{ p \}} \Pro_V \Pro_i 
 \Pro_{inner} \Pro_\pi \Pro_{J,\lambda} \ket{\Nj,\{ p \}} = 
 \frac{1}{\Omega} \sum_{\{ p \}} \bra{ \Nj,\{ p \}} \Pro_V \Pro_i \ket{\Nj,\{ p \}}
\end{eqnarray}
where we have used, in the last equality, the factorization (\ref{factor2}) and 
the idempotency of projectors $\Pro_{inner},\Pro_\pi,\Pro_{J,\lambda}$. Finally, 
the operators $\Pro_V$ and $\Pro_i$ can be swapped in position using again the 
factorization (\ref{factor2}), the commutation between $\Pro_{inner}$ and all 
other projectors, the ciclicity of $\Pro_\pi,\Pro_{J,\lambda}$, and the fact that 
$\ket{\Nj,\{ p \}}$ are eigenstates of total energy-momentum. Hence, the relative probability 
of a channel $\Nj$ is proportional its microcanonical weight, defined as:
\begin{equation}\label{microch1}
 p_{\Nj} \propto \Omega_{\Nj} \equiv 
 \sum_{\{ p \}} \bra{ \Nj,\{ p \}} \Pro_i \Pro_V \ket{\Nj,\{ p \}} 
\end{equation}

The microcanonical channel weight $\Omega_{\Nj}$ has been calculated explicitely
in ref.~\cite{bf4} for an ideal relativistic gas of particles with spin with the 
full Poincar\'e projector (\ref{poincare}) in a quantum field theoretical framework. 
The obtained expression is essentially the same as the one would get in a multiparticle 
approach, i.e. working in the multiparticle tensor space with symmetrization for
bosons and antisymmetrization for fermions; the only relevant quantum field effect 
being is an overall immaterial factor $\bra{0} \Pro_V \ket{0}$. Let 
$N$ be the total number of particles in the channel, i.e. $\sum_{j=1}^K N_j=N$; $S_j$ 
and $\eta_j$ respectively the spin and the intrinsic parity of the $j$-th particle 
species, $p_n$ the four-momentum of the n-th particle.
Then, for a spherical cluster, the microcanonical channel weight reads \cite{bf4}:
\footnote{We take this opportunity to notice that in the formula (82) in ref.~\bf{4} 
there was a factor 1/2 in excess.} 
\begin{eqnarray}\label{microch2}
&&\Omega_{\Nj}=  \bra{0} \Pro_V \ket{0} 
 \sum_{ \rhov  }\left[ \prod_{j=1}^{K}  \frac{ \chi(\rho_j) ^{b_j}}{N_j!} 
\right]\; \frac{1}{4\pi} \int_{0}^{4 \pi} \d \psi \; 
 \left[\prod_{j=1}^{K} \prod_{\parti_j=1}^{N_j} \int \d^3 {\rm p}_{\parti_j} \right]  
\\ \nonumber
&\times& \delta^4 \left(P - \sum_{\parti=1}^{N} p_\parti \right) \sin \frac{\psi}{2} \; 
\sin\left[\left(J+\frac{1}{2} \right) \psi \right] \prod_{j=1}^{K}\left[\prod_{\parti_j=1}^{N_j} 
\left[ \frac{\sin[(S_j+\frac{1}{2}) 
{\parti_j}\psi]}{\sin(\frac{\parti_j\psi}{2})} \right]^{h_{\parti_j}(\rho_j)} \right]  
\\ \nonumber
&\times&  \left(\prod_{j=1}^{K} \prod_{\parti_j=1}^{N_j} 
F^{(s)}_V({\bf p}_{\rho_j(\parti_j)}-{\sf R}_{3}^{-1}(\psi){\bf p}_{\parti_j}) + 
\Pi \Pi_f \prod_{j=1}^{K}\prod_{\parti_j=1}^{N_j} 
F^{(s)}_V({\bf p}_{\rho_j(\parti_j)}+{\sf R}_{3}^{-1}(\psi){\bf p}_{\parti_j})\right) 
\end{eqnarray}
where 
\begin{equation}
\label{giunta3b}
\Pi_f=\prod_{j=1}^{K} \eta_j^{N_j}
\end{equation}
and $\rhov = (\rho_1, \ldots, \rho_k)$ is a set of permutations, $\rho_j$ belonging 
to the permutation group $S_{N_j}$; $\chi(\rho_j)$ is the parity of the $j$-th 
permutation and $b_j=0,1$ if the species $j$ is a boson or a fermion respectively; the 
symbol $h_{\parti_j}(\rho_j)$ in (\ref{microch2}) stands for the number of cyclic
permutation with $\parti_j$ elements in $\rho_j$ so that $\sum_{\parti_j=1}^{\infty} 
\parti_j h_{\parti_j}(\rho_j)=N_j$~\footnote{The set of integers $h_1,\ldots,h_{N} 
\equiv \{h_{n}\}$, is usually defined as a {\em partition} of the integer $N$ in the 
multiplicity representation.}. In eq.~(\ref{microch2}), $F^{(s)}_V$'s are Fourier 
integrals over a spherically symmetric volume, which for a sharp sphere read:
\begin{eqnarray}\label{fourier}
&&F^{(\circ)}_V({\bf p}_{\rho(\parti)}-{\sf R}_{3}^{-1}(\psi){\bf p}_\parti) 
= \frac{1}{(2\pi)^3} \int_V \d^3 {\rm x} \; 
\e^{i {\bf x \cdot}({\bf p}_{\rho(\parti)}-{\sf R}_{3}^{-1}(\psi){\bf p}_\parti)} 
\\ \nonumber 
&&=\frac{ R^2}{2 \pi^2}
\frac{j_1(| {\bf p}_{\rho(\parti)}-{\sf R}_{3}^{-1}(\psi){\bf p}_\parti |R)}
{| {\bf p}_{\rho(\parti)}-{\sf R}_{3}^{-1}(\psi){\bf p}_\parti |} 
\end{eqnarray}
$R$ being the radius, $j_1$ the spherical Bessel function of the 
first kind and ${\sf R}_{3}(\psi)$ is a rotation of an angle $\psi$ along the $z$ axis.
The factor $\bra{0} \Pro_V \ket{0}$, as has been mentioned, is immaterial as it
cancels out in the ratios between different channels.

In the eq.~(\ref{microch2}) the dependence on the cluster polarization state $\lambda$ 
has disappeared because of spherical simmetry \cite{bf4}. 

If, in eq.~(\ref{microch2}), we sum up over all angular momenta $J$ and neglect 
all permutations except the identity (corresponding to the Boltzmann statistics), 
we obtain the more familiar expression:
\begin{equation}\label{microchapp}
\Omega_{\Nj}=  \bra{0} \Pro_V \ket{0} \frac{V^N}{(2\pi)^3N}
\left[ \prod_{j=1}^{K} \frac{(2S_j+1)^N_j}{N_j!} \right] \int \d^3 {\rm p}_1 
 \ldots \int \d^3 {\rm p}_N  \delta^4 \left(P - \sum_{\parti=1}^{N} p_\parti 
 \right) 
\end{equation}
which can be used to show that the dynamical matrix element in the cluster's
decay, according to the SHM, is proportional to $P \cdot p_i/rho$ for each particle, 
$\rho$ is the proper energy density of the cluster \cite{review}.

\subsection{Internal symmetries}
\label{internal}

The eq.~(\ref{microch2}) only contains the Poincar\'e group projector (\ref{poincare});
we now have to include the internal symmetry projector $\Pro_{inner}$. First of
all, we will disregard altogether the projector ${\sf P}_\Qf$ on the abelian charges 
baryon number and strangeness, as this gives rise to non-trivial coefficients 
(i.e. 0 or 1) and can be easily implemented algorithmically just by enforcing 
$\sum_{j} {\bf q}_j N_j=\Qf$, where ${\bf q}_j$ is the vector of abelian charges 
for the species $j$. 

As has been mentioned, we will work in the multiparticle tensor space instead
of Fock space and, for this purpose, it is convenient to introduce the concept of 
particle {\em type}. Particles of the same type are to be taken as identical, 
yet in a different charge state. We will take as identical (hence of the same type)
light-flavoured non-strange mesons belonging to the same isospin multiplet {\em or} 
if they are a particle-antiparticle pair. For instance, $\pi^+$,$\pi^-$, and $\pi^0$ 
belong to the same type, that one can define as the pion. Similarly, p and 
$\bar {\rm p}$ or K$^+$ and K$^-$ belong to the same type, while p and n, or
K$^+$ and K$^-$ do not because, albeit forming an isodoublet, are not light-flavoured 
mesons.

Let us denote with $L_j$ the number of particles of the type $j$ in a given channel 
$\Nj$ (a channel $\Nj$ completely defines the corresponding set $\{L_j\}$, while the 
converse is not true). If $\rho_j \in S_{L_j}$ is a permutation 
of the integers $1,\ldots, L_j$, $\chi(\rho_j)$ its parity and $b_j=0$ or $b_j=1$ 
if particles of type $j$ are bosons or fermions respectively, the general final state 
$\ket{f}$ in the multiparticle tensor space can be written factorizing groups of 
particles of the same type as:
\begin{equation}\label{statof1}
\ket{f} = \sum_{ \rhov} \left[ \prod_{j=1}^{K} \frac{ \chi(\rho_j)^{b_j}}{\sqrt{L_j !}} 
 \right] \prod_{j=1}^{K}\ket{ I_j ,\{k_{ \rho_{j}(l_j) }  \} , \{ I_3^{\rho_{j}(l_j)}\} , 
\{\eta_{\rho_{j}(l_j)}\} , \{{\bf q}_{\rho_{j}(l_j)}\}  } \; .
\end{equation}
$K$ being now the total number of types; $\rhov$ the set of permutations $\rho_1,\ldots,
\rho_K$, $k_{l_j}$ the kinematical variables of the particle $l_j$ (momentum and 
polarization), $\eta_{l_j}$ its parity and ${\bf q}_{l_j}$ its quantum numbers; 
$I_3^{l_j}$ is the isospin third component of the $l_j$-th particle of the type $j$. 
If we define:
\begin{equation}\label{statof1aa}
\ket{f_{\rhov}} \equiv \prod_{j=1}^{K}\ket{I_j ,
\{ k_{ \rho_{j}(l_j) }  \} , \{ I_3^{\rho_{j}(l_j)}\} , 
\{\eta_{\rho_{j}(l_j)}\} , \{{\bf q}_{\rho_{j}(l_j)}\}  } 
\end{equation}
then $\ket{f}$ reads, according to (\ref{statof1}):
\begin{equation}\label{statof1aaa}
\ket{f} \equiv \sum_{\rhov} \left[ \prod_{j=1}^{K} \frac{\chi(\rho_j)^{b_j}}
{\sqrt{L_j !}}\right] \; \ket{f_{\rhov}} \;
\end{equation}
and the {\em state weight} $\omega_f \equiv \bra{f}\Pro_i \Pro_V \ket{f}$ for a 
channel $\Nj$ in the multiparticle tensor space can then be written in terms of 
the corresponding set $\{L_j\}$ as:
\begin{equation}\label{eqn:trm7adummy4}
  \omega_f= \sum_{\sigmav} \sum_{\rhov} \left[ \prod_{j=1}^{K} 
  \frac{\chi(\rho_j \sigma_j)^{b_j}}{L_j!} \right] \bra{f_{\sigmav}}\Pro_i \Pro_V 
  \ket{f_{\rhov}} =\sum_{ \rhov } \left[ \prod_{j=1}^{K} \chi(\rho_j)^{b_j} \right] 
  \bra{f_{\iotav}} \Pro_i \Pro_V \ket{f_{\rhov}} \; . 
\end{equation}
where $\iotav$ is the identical permutation. 

We shall start studying the action of projectors on states with definite $C$-parity
$\chi$, $I$ and $I_3$ bearing in mind that relevant projectors can be moved to the 
right of $\Pro_V$ to act on $\ket{f}$. As already stated, the projector $\Pro_C$ is 
meaningful only if the cluster is completely neutral and reads: 
\begin{equation}\label{eqn:isoscp1}
\Pro_{C} \ket{f} = \frac{{\sf I} + \chi \widehat{{\sf C}}}{2}\ket{f} 
\end{equation} 
where $\widehat{{\sf C}}$ is the charge-conjugation operator transforming the state 
$\ket{f}$ into:
\begin{equation}\label{eqn:isoscp1a}
 \widehat{{\sf C}}\ket{f} = \sum_{\rhov} \left[ \prod_{j=1}^{K} 
 \frac{ \chi(\rho_j)^{b_j}}{\sqrt{L_j !}} \right] \; \widehat{{\sf C}}\ket{f_{\rhov}}
 = \chi_C \sum_{ \rhov } \left[ \prod_{j=1}^{K} \frac{ \chi(\rho_j)^{b_j}}{\sqrt{L_j!}} 
 \right]\; \ket{ \overline{f}_{\rhov}}
\end{equation} 
In the above equation:
\begin{equation}\label{eqn:isoscp1abis}
 \ket{\overline{f}_{\rhov}} =  \prod_{j=1}^{K}\ket{ I_j ,
 \{ k_{ \rho_{j}(l_j) }  \} , \{ -I_3^{\rho_{j}(l_j)}\} , 
 \{\overline\eta_{\rho_{j}(l_j)}\} , \{-{\bf q}_{\rho_{j}(l_j)}\}}  
\end{equation} 
and $\chi_C$ is the product of intrinsic C-parities of the completely neutral mesons 
and the charge conjugation phase factors of non-strange charged light-flavoured mesons,
defined in Appendix A. The symbol $\overline{\eta}_{l_j}$ stands for the parity of the 
charge-conjugated $l_j$-th particle. In order to ensure the commutation between $\Pro_C$ 
and the space reflection operator $\widehat {\sf \Pi}$, which has already been used 
in this Section, we have to set $\overline\eta = \eta$ for all particles. This is 
obvious for bosons but not for fermions, whose parity are arbitrary provided that 
$|\eta| = 1$ and $\eta\overline\eta=-1$. We then set 
$\eta_{{\rm B}}=\overline{\eta}_{{\rm B}}=i$ for baryons in order to meet all requirements. 

Let us now move to the isospin projector which can be written formally as:
\begin{equation}\label{eqn:isoscp4}
{\sf P}_{II_3}= \ket {I,I_3 }\bra{ I,I_3 } \; .
\end{equation}  
Preliminarly, it is useful to write the state $\ket{f_{\rhov}}$ as the tensor
product of a ket $\ket{s}$ including kinematical variables $k$ and parities $\eta$, 
and an internal part:
\begin{equation}\label{eqn:statof1ab}
 \ket{f_{\rhov}}=\ket{s_{\rhov}}\prod_{j=1}^{K} \ket{I_j
,\{ I_3^{\rho_{j}(l_j)}\}, \{{\bf q}_{\rho_{j}(l_j)}\} } \equiv \ket{s_{\rhov}}
 \prod_{j=1}^{K} \ket{I_j,\{ I_3^{\rho_{j}(l_j)}\}, \{{\bf q}_{\rho_{j}(l_j)}\} }
\end{equation}
so that:
\begin{equation}\label{eqn:statof1a}
 \ket{ f} \equiv \sum_{\rhov} \left[ \prod_{j=1}^{K} 
 \frac{ \chi(\rho_j)^{b_j}}{\sqrt{L_j !}} \right] \; \ket{s_{\rhov}} 
 \prod_{j=1}^{K}\ket{ I_j,\{ I_3^{\rho_{j}(l_j)}\}, \{{\bf q}_{\rho_{j}(l_j)}\}}
\end{equation}
Now, by using eqs.~(\ref{eqn:isoscp1}) and (\ref{eqn:isoscp1a}), the state weight 
in eq.~(\ref{eqn:trm7adummy4}) can be expanded. Denoting the isospin projection 
coefficients by:
\begin{eqnarray}\label{eqn:isosimbols}
&&\mathcal{I}_{\rhov}^{\Nj}\left( I, I_3\right) \equiv \left[ \prod_{j=1}^{K} 
\bra{ I_j, \{ I_3^{l_j}\}  
} \right] \ket{ I,I_3 } \bra{ I,I_3 } \left[ \prod_{j=1}^{K} \ket{  \{
I_j, I_3^{\rho_{j}(l_j)}\} } \right] \qquad \\ \nonumber
&&\overline{\mathcal{I}}_{\rhov}^{\Nj}\left(I,I_3 \right) \equiv 
\left[ \prod_{j=1}^{K} \bra{ I_j, \{ I_3^{l_j}\}} \right] 
 \ket{ I,I_3 } \bra{ I,I_3 } \left[ \prod_{j=1}^{K} 
 \ket{I_j,\{-I_3^{\rho_{j}(l_j)}\}}\right] 
\end{eqnarray}
the state weight turns into, by using the factorization of projectors in 
eq.~(\ref{factor1}):
\begin{eqnarray}\label{sweight}
 \omega_f &=&\frac{1}{2}  
  \sum_{\rhov }\left[ \prod_{j=1}^{K}  \chi(\rho_j)^{b_j} \right] \left( 
  \mathcal{I}_{\rhov}^{\Nj}\left( I, I_3\right)
  \bra{s_{\iotav}} {\sf P}_{P J \lambda \Pi} \Pro_V \ket{s_{\rhov}} \; 
  \prod_{j=1}^{K}\prod_{l_j=1}^{L_j}\delta_{{\bf q}_{\rho_{j}(l_j)}\;{\bf q}_{l_j}} 
 \right.\nonumber  \\ 
&+& \left.C \chi_{C}  
 \overline{\mathcal{I}}_{\rhov}^{\Nj}\left(I,I_3 \right)
 \bra{s_{\iotav}} {\sf P}_{P J \lambda \Pi} \Pro_V \ket{\overline{s}_{\rhov}} \; 
  \prod_{j=1}^{K}\prod_{l_j=1}^{L_j}\delta_{-{\bf q}_{\rho_{j}(l_j)}\;{\bf q}_{l_j}} \right)
\end{eqnarray}
where Kronecker factors $\delta_{-{\bf q}_{\rho_{j}(l_j)}}^{{\bf q}_{l_j}}$ and 
$\delta_{{\bf q}_{\rho_{j}(l_j)}}^{{\bf q}_{l_j}}$ stem from the vanishing of scalar 
products between single particle states with different baryon number and strangeness. 
Note that, as the parity of particles and antiparticles are the same by construction:
\begin{equation}\label{defckin}
 \ket{\overline{s}_{\rhov}}\equiv  \ket{ \{ k_{\rho_{j}(l_j)}  \}, 
 \{\overline{\eta}_{\rho_{j}(l_j)}\}} =
 \ket{ \{ k_{\rho_{j}(l_j)}  \}, \{\eta_{\rho_{j}(l_j)}\}}= \ket{s_{\rhov}}
\end{equation}
therefore, taking into account eq.~(\ref{defckin}), we rewrite eq.~(\ref{sweight}) 
as:
\begin{eqnarray}\label{sweight2}
\omega_f &=&\frac{1}{2} \sum_{\rhov}\left[ \prod_{j=1}^{K}  \chi(\rho_j)^{b_j} \right] 
\bra{s_{\iotav}} {\sf P}_{P J \lambda \Pi} \Pro_V \ket{s_{\rhov}}  \\ \nonumber 
 &\times& \left( \mathcal{I}_{\rhov}^{\Nj}\left( I, I_3\right)\; 
\prod_{j=1}^{K}\prod_{l_j=1}^{L_j}\delta_{{\bf q}_{\rho_{j}(l_j)}\;{\bf q}_{l_j}} 
 + \chi \chi_{C} \overline{\mathcal{I}}_{\rhov}^{\Nj}\left(I,I_3 \right) \; 
 \prod_{j=1}^{K}\prod_{l_j=1}^{L_j}\delta_{-{\bf q}_{\rho_{j}(l_j)}\;{\bf q}_{l_j}} 
 \right) \; .
\end{eqnarray}

The microcanonical weight of a multi-hadronic channel can be then calculated (for 
spherical clusters) from (\ref{sweight2}) by summing over particle momenta $p$ and 
polarizations $\sigma$ and averaging over the initial polarization of the cluster. 
In formula:
\begin{equation}\label{integrate}
 \frac{1}{(2J+1)}\sum_{\lambda}\left[ \prod_{j=1}^{K}\sum_{\{\sigma_{l_j}\}}\right] 
 \left\{ \prod_{j=1}^{K} \frac{1}{N_j!} \left[\prod_{\parti_j=1}^{N_j} 
 \int \d^3 {\rm p}_{\parti_j}\right] \right\}
\end{equation}
where $K$ is the number of types, $N_j$ is the number of particles of the type $j$ 
and the factor $\prod_{j=1}^{K}1/N_j!$ is needed in order avoid multiple 
counting of (anti)symmetric basis tensors when integrating over all particle momenta.

The integration over kinematical variables, understood in $\ket{s_{\rhov}}$, gives 
rise to the same expression as in (\ref{microch2}), with the difference that
now $j$ labeling types and not species. There is also an additional coefficient 
related to internal symmetry (isospin and C-parity):
\begin{eqnarray}\label{microchf}
&&\Omega_{\Nj}=  \bra{0} \Pro_V \ket{0} 
 \sum_{ \rhov  } \left[ \prod_{j=1}^{K} \chi(\rho_j)^{b_j} \right]\; 
 \frac{1}{8\pi} \int_{0}^{4 \pi} \d \psi \; \left[\prod_{j=1}^{K}\frac{1}{N_j!} 
 \prod_{\parti_j=1}^{N_j} \int \d^3 {\rm p}_{\parti_j} \right] \\ \nonumber
&&\times \delta^4 \left(P - \sum_{\parti=1}^{N} p_\parti \right) \sin \frac{\psi}{2} \; 
 \sin\left[\left(J+\frac{1}{2} \right) \psi \right] \prod_{j=1}^{K} \left[\prod_{l_j=1}^{L_j} 
 \left[ \frac{\sin[(S_j+\frac{1}{2}){l_j}\psi]}
 {\sin(\frac{l_j\psi}{2})}\right]^{h_{l_j}(\rho_j)}\right] \\ \nonumber
&&\times \left( \prod_{j=1}^{K} \prod_{l_j=1}^{L_j} 
 F^{(s)}_V({\bf p}_{\rho_j(l_j)}-{\sf R}_{3}^{-1}(\psi){\bf p}_{l_j}) + \Pi \Pi_f 
 \prod_{j=1}^{K}\prod_{l_j=1}^{L_j} 
 F^{(s)}_V({\bf p}_{\rho_j(l_j)}+{\sf R}_{3}^{-1}(\psi){\bf p}_{l_j})\right) \\ \nonumber
&&\times \left( \mathcal{I}_{\rhov}^{\Nj}\left( I, I_3\right)\; 
 \prod_{j=1}^{K}\prod_{l_j=1}^{L_j}\delta_{{\bf q}_{\rho_{j}(l_j)} {\bf q}_{l_j}} 
 +\chi \chi_{C} \overline{\mathcal{I}}_{\rhov}^{\Nj}\left(I,I_3 \right) \; 
\prod_{j=1}^{K}\prod_{l_j=1}^{L_j}\delta_{-{\bf q}_{\rho_{j}(l_j)} {\bf q}_{l_j}} \right)
\end{eqnarray}

The eq.~(\ref{microchf}) is the final expression of the microcanonical channel
weight for a multi-hadronic channel. This applies to completely neutral clusters 
with $C$-parity $\chi$. For charged clusters the microcanonical channel weight can be 
obtained by removing the C-parity projector, i.e. setting $\chi=0$ in the above 
equation and multiplying the result by 2.

\section{Numerical computation}\label{numerics}

The channel weight in eq.~(\ref{microchf}) is the basic expression to calculate 
exclusive channel rates; it cannot be worked out analitically, but it can be 
evaluated numerically. According to eq.~(\ref{microchf}), the task is indeed twofold:
first, the computation of the isospin coefficients (see eq.~(\ref{eqn:isosimbols}) 
and, second, that of multi-dimensional momentum integrals. The sum over permutations
can be made with well known algorithms. In order to compute the isospin coefficient 
we have designed a suitable recursive algorithm which is described in Appendix B, 
while in this section we focus on the problem of computing the momentum integrals. 

For properly relativistic particles this problem is known not to have an analytic 
solution. Previous attempts to obtain sufficiently accurate estimates \cite{cerang4} 
resorted to Monte-Carlo integration and this is the technique we take.
For the sake of simplicity, we will describe our method for a channel with particles 
of different species; the generalization to identical particles entails a sum over
permutations in the integrand function and does not involve special difficulties. 
In this case, for a cluster at rest, the general momentum integral in 
eq.~(\ref{microchf}) can be written as:
\begin{equation}\label{nummed1}  
 \int \d^3 {\rm p}_{1} \ldots \d^3 {\rm p}_{N}  \delta \left(M - \sum_{\parti=1}^{N} 
  \varepsilon_\parti \right)\delta^3 \left(\sum_{\parti=1}^{N} {\bf p}_\parti \right) 
 W(\{{\bf p}_\parti \}) 
\end{equation}
where the function $W$ is, up to a constant:
\begin{eqnarray}\label{functionW}
  W(\{{\bf p}_\parti \})&=&\int_{0}^{4 \pi} \d \psi \; \sin \frac{\psi}{2} \; 
  \sin\left[\left(J+\frac{1}{2} \right) \psi \right] \prod_{\parti=1}^{N}  
\left[ \frac{\sin[(S_\parti+\frac{1}{2})\psi]}{\sin(\frac{\psi}{2})} \right] \\ \nonumber
 &\times& \left( \prod_{\parti=1}^{N}  
 F^{(s)}_V({\bf p}_{\parti}-{\sf R}_{3}^{-1}(\psi){\bf p}_{\parti}) + 
 \Pi \Pi_f \prod_{\parti=1}^{N} F^{(s)}_V({\bf p}_{\parti}+
 {\sf R}_{3}^{-1}(\psi){\bf p}_{\parti})\right) \; . 
\end{eqnarray}
In eq.~(\ref{nummed1}) the last momentum variable ${\rm p}_N$ can be integrated 
away at once with the $\delta^3$ of momentum conservation, yielding:
\begin{equation}\label{nummed3}
 \int \d^3 {\rm p}_{1} \ldots \d^3 {\rm p}_{N-1}  \\ \nonumber 
 \delta \left(M - \sum_{\parti=1}^{N-2} \varepsilon_\parti - \sqrt{{\rm p}^2_{N-1}
 +m^2_{N-1}}-\sqrt{| -{\bf P}_{N-2}-{\bf p}_{N-1} |^2+m^2_N} \right) 
 W(\{{\bf p}_\parti \} ) 
\end{equation}
where: 
\begin{displaymath}
{\bf P}_{N-2}\equiv \sum_{\parti=1}^{N-2} {\bf p }_\parti \; .
\end{displaymath}
Note that in eq.~(\ref{nummed3}), for simplicity, we have used the same symbol for 
$W(\{{\bf p}_\parti \})$ keeping in mind that indeed, after the integration, this
is a different function as it only depends on the set of momenta ${\bf p}_1,\ldots,
{\bf p}_{N-1}$. Denoting with ${\rm p}_{N-1}^{(A)}$ and ${\rm p}_{N-1}^{(B)}$
the two zeroes of the argument of the $\delta$ in eq.~(\ref{nummed3}), this can be
rewritten as:
\begin{eqnarray}\label{nummed4}
&& \int \d^3 {\rm p}_{1} \ldots \d^3 {\rm p}_{N-1} 
 \left| \frac{\varepsilon_N\varepsilon_{N-1}}{{\rm p}_{N-1}(\varepsilon_{N-1}+
  \varepsilon_{N})+\varepsilon_{N-1}{\bf P}_{N-2}\cdot \hat{{\bf p}}_{N-1}}\right| 
  \nonumber \\  
&&\times  \left[ \delta({\rm p}_{N-1}-{\rm p}_{N-1}^{(A)})+
 \delta({\rm p}_{N-1}-{\rm p}_{N-1}^{(B)})\right] W(\{{\bf p}_\parti \}) 
\end{eqnarray}
where $\hat{{\bf p}}_{N-1}$ is the versor of ${\bf p}_{N-1}$ and ${\rm p}_{N-1}^{(A)}$ 
and ${\rm p}_{N-1}^{(B)}$ fulfill the following equation:
\begin{eqnarray}
\label{nummed5}
&& \!\!\!\!\! \left[ \left(\hat{{\bf p}}_{N-1} \cdot {\bf P}_{N-2} \right)^2-
\left(M- \sum_{\parti=1}^{N-2} \varepsilon_\parti\right)^2\right] {\rm p}^2_{N-1}
+\left(\hat{{\bf p}}_{N-1} \cdot {\bf P}_{N-2} \right) \left[{\rm P}^2_{N-2} - 
\left(M- \sum_{\parti=1}^{N-2} \varepsilon_\parti\right)^2+m^2_N-m^2_{N-1} \right]  
 {\rm p}_{N-1}  \nonumber \\ 
&& + \frac{1}{4} \left[{\rm P}^2_{N-2} - 
 \left(M- \sum_{\parti=1}^{N-2} \varepsilon_\parti\right)^2+m^2_N-m^2_{N-1} 
 \right]^2-m^2_{N-1} \left( M- \sum_{\parti=1}^{N-2} \varepsilon_\parti\right)^2 
 = 0 \;.
\end{eqnarray}
If we now let:
\begin{equation}\label{nummed4nuova2}
\Phi^A = \left| \frac{\varepsilon_N^{(A)}\varepsilon_{N-1}^{(A)}}
{{\rm p}_{N-1}^{(A)}\left(\varepsilon_{N-1}^{(A)}+\varepsilon_{N}^{(A)} \right)+
 \varepsilon_{N-1}^{(A)}{\bf P}_{N-2}\cdot \hat{{\bf p}}_{N-1}} \;\right|
\end{equation}
where:
\begin{equation*}
 \varepsilon_N^{(A)}=\sqrt{\left|{\rm p}_{N-1}^{(A)} \hat{{\bf p}}_{N-1}+ {\bf P}_{N-2} 
 \right|^2+m^2_N} \qquad {\rm and } \qquad \varepsilon_{N-1}^{(A)} =
 \sqrt{\left({\rm p}_{N-1}^{(A)} \right)^2+m^2_{N-1}} \; \;
\end{equation*}
and similarly for ${\rm p}_{N-1}^{(B)}$, the eq.~(\ref{nummed4}) can be finally 
written as: 
\begin{equation}\label{nummed4nuova}
\int \d^3 {\rm p}_{1} \ldots {\rm p}_{N-2}\, \d \Omega_{{N-1}} 
 \left[ \Phi^A W(\{{\bf p}_\parti \}^A) + \Phi^B W(\{{\bf p}_\parti \}^B)\right] 
\end{equation}
where $\Omega_{{N-1}}$ is the solid angle of $\hat{{\bf p}}_{N-1}$ and 
$W(\{{\bf p}_\parti \}^A)$ stands for the function $W$ in eq.~(\ref{functionW}) 
evaluated on the momenta:
\begin{displaymath}
\{{\bf p}_\parti \}^A=\left({\bf p}_1, \ldots,{\bf p}_{N-2},{\rm p}_{N-1}^{(A)}
 \hat{{\bf p}}_{N-1},-\left( {\rm p}_{N-1}^{(A)} \hat{{\bf p}}_{N-1}+{\bf P}_{N-2} 
 \right) \right) \; ;  
\end{displaymath}
and similarly for $W(\{{\bf p}_\parti \}^B)$.

In order to calculate eq.~(\ref{nummed4nuova}) we have to perform a $3N-4$-dimensional 
momentum integration plus one further integration over the angle $\psi$ hidden
in the function $W$. Overall, this is a $3N-3$ dimensional integration which is carried 
out by using an importance sampling Monte-Carlo method, designed to achieve the 
best performance. 

\subsection{Importance Sampling}\label{sampling}

The importance sampling method is a well-known method to perform Monte-Carlo 
integration. The idea is to find an auxiliary function $g(x)$ which is at the same 
time easy to sample and as similar as possible to the integrand function to maximize 
efficiency. Each random extraction in $x$ is then weighted by the ratio $f(x)/g(x)$ 
and an estimator of the integral is given, after $N$ extractions, by
\begin{equation*}
    \sum_{i=1}^N f(x_i)/g(x_i) 
\end{equation*}  
In our case, we have to extract $3N-3$ variables: $N-2$ momenta, $N-1$ solid angles, 
and one further angle $\psi$. We extract all angles according from a flat distribution
while for momenta, our method is based on the use of the asymptotic limit ($N \to 
\infty$) of the integrand; this is a known one, as for $N \to \infty$
the microcanonical $N$-body phase space should converge to its canonical ensemble
limit, which consists - in the Boltzmann limit - of a factor $\exp[-\varepsilon/T]$ 
for each particle. Therefore, for each particle, we expect:
\begin{equation}\label{ke}
 \p^2 \d\p \, \e^{-\varepsilon/T} \to \d t \sqrt{t(t+2m_j)}(t+m_j) \e^{-t/T} \; , 
\end{equation}
for its kinetic energy distribution. The temperature $T$ in eq.~(\ref{ke}) is a 
parameter to be chosen to minimize the difference between integrand and auxiliary
function. We have calculated it, along with a set of chemical potentials 
associated to each conserved charge (electric, baryonic and strange) by equating 
the total energy, momentum and charges in the microcanonical ensemble with their 
average value in the grand-canonical ensemble, which is precisely the saddle-point 
equation governing the asymptotic expansion of the microcanonical partition 
function \cite{bf1}:
\begin{eqnarray}\label{tmu}
 M&=&T^2 \frac{\partial}{\partial T} \sum_{j}z_j(T) \e^{\muv \cdot {\bf q}_j/T} 
 \\ \nonumber
 {\bf Q}&=&\sum_{j}{\bf q}_jz_j(T)\e^{\muv \cdot {\bf q}_j/T}
\end{eqnarray}
where $M$ is the cluster mass and ${\bf q}_j$ is the set of abelian charges of 
the species $j$; $\muv$ is a set of chemical potentials and:
\begin{equation}\label{tmu2}
z_j(T) = (2S_j+1) \frac{V}{(2\pi)^3} \int \d^3 {\rm p} 
\exp{\left(-\varepsilon_j/T\right)}=\frac{(2S_j+1)V}{2 \pi^2} m^2_j 
 T {\rm K}_2\left(\frac{m_j}{T}\right)
\end{equation}
where ${\rm K}_2(m_j/T)$ is the modified Bessel function of the $2^{{\rm nd}}$
kind.

Unfortunately, the function (\ref{ke}) is not a practical auxiliary function
as it cannot be sampled fast enough. Its integral cannot be inverted analytically 
and there are not optimized sampling algorithms either in the ultra-relativistic 
or non-relativistic limit where it can be approximated as:
\begin{eqnarray}\label{tmu4}
&&t \gg m_j \qquad \rightarrow \qquad \sim A(t^2\;\e^{-t/T}) \\ \nonumber 
&&t \ll m_j \qquad \rightarrow \qquad \sim A(\sqrt{t}\;\e^{-t/T})
\end{eqnarray}
where $A$ is a normalization constant. We have then replaced the function (\ref{ke})
with the mathematical $\beta$ function:
\begin{equation}\label{betaf}
\beta(x)=\frac{\Gamma(a+b)}{\Gamma(a)\Gamma(b)}x^{a-1}(1-x)^{b-1} \qquad a,b>0 
 \qquad 0<x<1
\end{equation}
which can be very efficiently sampled \cite{tesigabb} with a fast rejection algorithm
\cite{ChengBB4}. In our case, $x$ is set to be the ratio $t/t_{{\rm max}}$, where 
$t$ is the kinetic energy of each particle and $t_{{\rm max}}$ is the highest available
kinetic energy, that is $M - \sum_{i=1}^N m_i$:
\begin{equation}\label{betaf2}
\beta \left(\frac{t}{t_{{\rm max}}} \right)=
\frac{\Gamma(a+b)}{\Gamma(a)\Gamma(b)}\left( \frac{t}{t_{{\rm max}}}\right)^{a-1}
\left(1- \frac{t}{t_{{\rm max}}}\right)^{t_{{\rm max}}/T}
\end{equation}
The constant $b$ is chosen to be $b=1+t_{{\rm max}}/T$ where $T$ is obtained from
solving eqs.~(\ref{tmu}). This choice is dictated by the fact that in the limit 
$t_{{\rm max}}/T \gg 1$ this distribution is very close to an exponential, like
in eq.~(\ref{ke}):
\begin{equation*}
 \lim_{\stackrel{t_{{\rm max}}/T \rightarrow \infty}{t \ll t_{{\rm max}} }} 
 \left( 1- \frac{t}{t_{{\rm max}}}\right)^{t_{{\rm max}}/T} \sim \; \e^{-t/T} \; .
\end{equation*}
The only problem of the distribution (\ref{betaf2}) is that it does not match properly
the actual one when $t$ is large, i.e. when $t \sim t_{{\rm max}}$; indeed, 
eq.~(\ref{betaf2}) goes to zero too rapidly and this makes the algorithm less 
effective in sampling the region in multi-dimensional momentum space where one 
particle in the channel carries most of the available kinetic energy. However,
such contributions are relevant only for channels with few light particles and
one much more massive. 

As has been shown in~\cite{tesigabb}, a good choice for the parameter $a$ is:
\begin{equation}
\label{tmu5}
a-1=\frac{1}{2}+\frac{3}{2} \; \e^{-2m_j}
\end{equation}
where $m_j$ is in GeV. The empirical dependence of $a$ on particle mass is such
that $a$ takes the values $3$ and $1.5$ respectively in the ultrarelativistic 
and non-relativistic limit, according to (\ref{tmu4}).

Summarizing, the steps of the Monte-Carlo integration algorithm are as follows:
\begin{itemize}
\item{} extraction of the angle $\psi$ with a flat distribution;
\item{} extraction of $N-1$ solid angles with a flat distribution;
\item{} extraction of $N-2$ kinetic energies according to $\beta$ distribution;
\item{} evaluation of the modulus $p_{N-1}$ solving eq.~(\ref{nummed5});
\item{} evaluation of ${\bf p}_{N}=-\sum_{\parti=1}^{N-1} {\bf p}_{\parti}$.
\end{itemize}

We note that eq.~(\ref{nummed5}) can have either one real solution or two real
solutions or none. In case of two solutions the integrand is evaluated on both
and averaged, while for no solutions the whole extraction procedure is repeated.
We also take into account, for channels with resonances, their mass broadening.  
In fact, masses are also randomly extracted, at each step of the Monte-Carlo 
integration, according to a relativistic Breit-Wigner distribution:
\begin{equation}\label{breitw}
  B_r(m) \equiv \frac{1}{2\pi}\frac{\Gamma_r}{(m-m_r)^2 + \Gamma_r^2/4}
\end{equation}
where $m_r$ is the central mass value and $\Gamma_r$ the width.

\subsection{Resonance decay channels}

\begin{figure}[htpb]
\begin{center}
\includegraphics[width=1.0\textwidth]{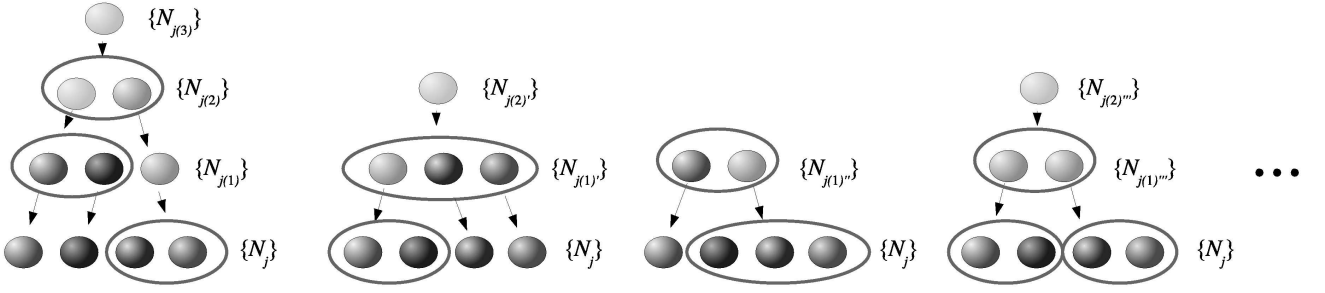}
\caption{\small{Examples of possible decay trees for a four particles channel. 
Circles encompass decay products of the particle at higher level.}}
\label{decay}
\end{center}
\end{figure}

As has been discussed at the end of Sect.~\ref{model}, for each exclusive channel
we assume the hadron-resonance gas picture, which prescribes that the probability 
of an exclusive channel with some set of final hadrons $\Nj$ is the sum, with suitable
weights, of the probability of all possible channels with hadrons and resonances
decaying - in a single or multiple steps - into the set $\Nj$. The weights are given
by products of branching ratios of parent hadrons and resonances. 
Therefore, given a final channel $\Nj$, the first task is to find all channels
which may have generated it, knowing the decay modes of all hadrons and resonances. 
The search of all parent channels is a multi-step recursive problem, in that many 
generation steps can occur. If we denote by $\Nj_{(1)}$ a channel which can 
directly decay into the channel $\Nj$, by $\Nj_{(2)}$ a channel which can directly 
decay in $\Nj_{(1)}$ and so on, one has to find all possible decay trees like those 
shown in fig.~\ref{decay}. In view of the large number of resonances, this task 
is a hard combinatorial problem and a suitable recursive algorithm has been designed to 
solve it.

\begin{figure}[htpb]
\begin{center}
\includegraphics[width=1.0\textwidth]{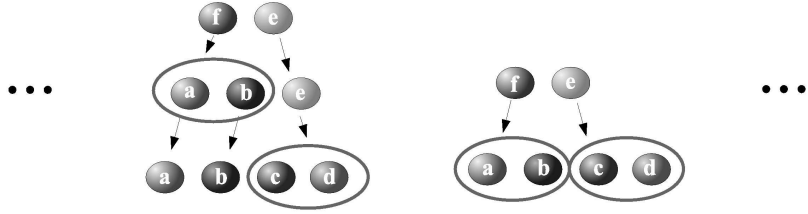}
\caption{\small{Examples of possible equivalent decay trees for a four particles 
channel.}}
\label{decay2}
\end{center}
\end{figure}

Our method is to list all subsets of particles in the channel $\Nj$ and check 
whether there is one or more resonances decaying in each subset. For instance, for 
a channel with four particles labelled with $1,2,3,4$, possible subsets are:
\begin{eqnarray*}
&&(1)(2,3,4)\qquad \qquad(2,3)(1)(4)\\ 
&&(2)(1,3,4)\qquad \qquad(2,4)(1)(3)\\ 
&&(3)(1,2,4)\qquad \qquad(3,4)(1)(2)\\ 
&&(4)(1,2,3)\qquad \qquad(1,2)(3,4)\\ 
&&(1,2)(3)(4)\qquad\qquad (1,3)(2,4)\\
&&(1,3)(2)(4)\qquad \qquad(1,4)(2,3)\\
&&(1,4)(2)(3)\qquad \qquad(1,2,3,4)\; .
\end{eqnarray*} 
Each resonance decaying into a subset gives rise to a possible upper-level parent 
channel made of the resonance replacing the given subset and the remaining hadrons
of the original channel. For each parent channel found, the procedure is iterated.
At the uppermost level, possibly one has a channel with only one massive resonance 
having the same quantum numbers as the initial state.

The complexity of this combinatorial problem is considerable. Actually, the number 
$B_n$ of subsets of a set of $n$ integers is known as the {\em Bell number} which 
is given in terms of a recurrence relation:
\begin{equation}
\label{eqn:bell}
B_{n+1}=\sum_{k=0}^{n} B_k {n \choose k} \; .
\end{equation}
This number grows quickly as a function of $n$, so that listing all possible parent 
channels becomes practically impossible for channels with more than $\sim 8$ 
particles ($B_{8}=4140$ and $B_{9}=21147$), also in view of the very large number
of resonances in the hadron spectrum. 

Indeed, this recursive search algorithm can give rise to multiple counting. This happens
if there are identical particles in the channel simply because some subsets are 
actually the same. For instance, if there are four particles, the subsets $(1,2)(3,4)$ 
and $(1,3)(2,4)$ are clearly equivalent if particles $2$ and $3$ are identical. This 
kind of double counting is quite easy to get round algorithmically {\it a priori}. 
Yet, double counting may occur even if particles in the channel are all different. 
Consider for instance two decay trees like those shown in fig.~\ref{decay2} with, on 
the left, the subset $(a)(b)(cd)$ and on the right the subset $(ab)(cd)$. In fact, both 
configurations can stem from the same channel $e,f$ at a different level, so this 
parent channel may appear twice in our parent channel search. This kind of multiple 
counting is avoided by re-checking {\it a posteriori} the full list of parent channels 
found.

Finally, the probability $\rho_{\Nj}$ of observing a final channel $\Nj$ can be written
as a finite sum over all parent channels:
\begin{eqnarray}\label{total}
\rho_{\Nj} &\propto& \tilde\Omega_{\Nj} \equiv \Omega_{\Nj}+{\rm BR}_{(1)}
\Omega_{\Nj_{(1)}}+{\rm BR}_{(2)}{\rm BR}_{(1)}\Omega_{\Nj_{(2)}}+ \ldots 
\nonumber \\
&+& {\rm BR}_{(1)'}\Omega_{\Nj_{(1)'}}+   
{\rm BR}_{(2)'}{\rm BR}_{(1)'}\Omega_{\Nj_{(2)'}}+ \ldots 
\end{eqnarray}
where ${\rm BR}_{(i)}$ is the product of branching ratios of particles in 
the channel $\Nj_{(i)}$ decaying into particles in the channel $\Nj_{(i-1)}$ 
and where $\tilde\Omega_{\Nj}$ is, by definition, the total channel weight, 
including contributions of parent channels.

\section{Analysis of \ee collisions at low energy}\label{ee}
 
As has been mentioned in the Introduction, the rates of exclusive hadronic channels 
can be measured only in low energy collisions (say $\lesssim 5$~GeV) because the 
large multiplicity of the final state at high energy makes a full identification
of particles impossible. There have been in the past some attempts to reproduce 
hadron multiplicities and some multi-pion(kaon) differential cross sections 
in low energy \ee collisions~\cite{eestat2} by using statistical-thermodinamical or 
statistical-inspired models. Yet, in none of those calculations the full set of 
conservation laws has been taken into account, because of the lack of a proper 
formulation of the microcanonical ensemble with intrinsic angular momentum and the 
involved numerical calculations. As we will show in Subsect.~\ref{testlaws}, this 
is a serious drawback because, when dealing with exclusive channels, all conservation 
laws play a major role. In fact, we now have all needed ingredients to make a proper 
test of the statistical hadronization model with exclusive channels: the statistical 
weight of multi-particle channels including exact energy-momentum and intrinsic 
angular momentum conservation \cite{bf4} (formula (\ref{microchf}) and a sufficient 
computing power. 

As discussed in Sect.~\ref{model}, at low energy the formation of a single cluster 
at rest in the centre-of-mass frame of an \ee collision is assumed. Its mass will 
therefore coincide with $\sqrt{s}$ and the other quantum numbers will be those of
the initial state. Particularly, in \ee collision, the hadron production is dominated 
by the diagram with an intermediate virtual photon (see fig.~\ref{eecoll}), 
so that the hadronizing cluster is assigned with a spin, parity and C-parity 
$J^{PC}=1^{--}$. On the other hand, isospin is not conserved in electromagnetic
interaction and it is therefore unknown; in the Vector Dominance Model (VDM) this 
depends on the coupling of the photon to different resonances, but we will be working 
in a mass region above 2 GeV, far from known resonance region (see discussion in the 
following). Therefore, we will assume an unknown statistical mixture of $I=0$ and 
$I=1$ initial state, neglecting interference terms, and introducing a free parameter 
${\rm I}_0$ such that for the mixed state:
\begin{displaymath}
{\rm I}_0\ket{0,0}\bra{0,0}+(1-{\rm I}_0) \ket{1,0}\bra{1,0},
\end{displaymath}

Finally, the geometry of the cluster needs to be fixed. We assume a spherical shape 
and a volume given by: 
\begin{equation}\label{endens}
  V = \frac{M}{\rho} = \frac{\sqrt s}{\rho}
\end{equation}
where $M$ is the mass and $\rho$ the energy density; this is taken to be a free 
parameter to be determined by comparing the model with the data.

Motivated by observations concerning hadron abundances at high energy, we allow 
deviation from the full statistical equilibrium of channels involving particles 
with strange valence quarks. This is done by introducing in the analysis the 
extra-strangeness suppression parameter $\gamma_S$ \cite{rafelski}. For its definition 
here to be in agreement with the formulae of inclusive multiplicities of hadrons 
in the canonical and grand-canonical ensembles, one just needs to multiply the 
microcanonical weight of a channel by $\gamma_S^{s_j}$, $s_j$ being the number of 
valence strange quarks of each particle:
\begin{equation}\label{strsupp}
\Omega_{\Nj} \rightarrow \left[ \prod_{j=1}^{K} 
 \left( \gamma_S^{s_j} \right)^{N_j}\right]\Omega_{\Nj}
\end{equation}
The $\gamma_S$ factor also applies to neutral mesons with hidden strange quark
content like $\eta$, $\phi$ etc. Since the wavefunction of such particles is in general 
a superposition like $C_{\uq} \uq\overline{\uq}+C_{\dq} \dq\overline{\dq}+C_{\sq}
\sq\overline{\sq}$ with $|C_{\uq}|^2+|C_{\dq}|^2+|C_{\sq}|^2=1$, only the 
component $\sq\overline{\sq}$ of the wavefunction is suppressed, i.e. we multiply
by:
\begin{displaymath}
\left[ \vert C_{\sq}\vert^2\gs^2+(1-\vert C_{\sq}\vert^2) \right] \; .
\end{displaymath} 
for each neutral meson. To calculate $\vert C_{\sq}\vert^2$, we have used mixing 
angles quoted by the Particle Data Book \cite{pdg2010}.

The branching ratios, masses and widths of hadrons and resonances needed to calculate
the exclusive channel rate according to formula (\ref{total}), have also been taken 
from the latest issue of the Particle Data Book \cite{pdg2010}. All hadrons up to a 
mass of 1.8 GeV for mesons and 1.9 for baryons have been included for the generation 
of parent channels. It is now appropriate to briefly discuss the possible contribution
of single resonance decay (off-peak) to the hadron production in \ee collisions at low energy.
In terms of the diagrammatic description in fig.~(\ref{decay}), these contributions 
correspond to the highest ancestor of the channel in the decay tree, being a single 
resonance with the same quantum numbers as the initial state. This contribution, if
relevant, cannot be subtracted away from the experimental data given the poor knowledge 
of resonances above 1.8-1.9 GeV mass. If one assumes that resonances can be identified
with clusters decaying statistically, then this contribution should be somehow taken
into account within the SHM calculation itself. On the other hand, if we refrain from
this assumption, we must move sufficiently far from the energy region where narrow 
resonances lie in order to minimize their impact on the cross section. Furthermore, 
we do not want to get over the charm quark production threshold and this constrains 
the energy interval to about 2-3 GeV. 

\begin{table}\label{fitsumm}
\begin{center}
\begin{tabular}{|c|c|c|c|c|c|c|}
\hline
$\sqrt{s}$~(GeV) & $\rho$~(GeV/fm$^3$) & $\gamma_{S}$ & $C_{\rho \gamma_{S}}$~(GeV/fm$^3$) 
& I$_{0}$ & A (nb GeV$^4$) & $\chi^2$/dof  \\
\hline
 $2.1$ & $0.24 \pm 0.17$ & $0.66 \pm 0.22$ & $0.011$ & $0.17 \pm 0.03$ & $0.028 \pm 0.002$  & 93.4/16 \\
\hline
 $2.2$ & $0.36 \pm 0.20$ & $0.86 \pm 0.22$ & $0.023$ & $0.15 \pm 0.03$ & $0.036 \pm 0.002$  & 82.6/14 \\
\hline
 $2.4$ & $0.44 \pm 0.30$ & $0.78 \pm 0.36$ & $0.024$ & $0.23 \pm 0.04$ & $0.017 \pm 0.001$  & 55.4/17 \\
\hline 
 $2.6$ & $0.56 \pm 0.36$ & $0.62 \pm 0.47$ & $-0.009$ & $0.53 \pm 0.07$ & $0.017 \pm 0.002$ & 44.9/12 \\
\hline 
\end{tabular}
\end{center}
\caption{Summary of the fit results to multi-hadronic exclusive channels at different
centre-of-mass energies. Also shown the correlation coefficient of $\rho$ and $\gamma_S$.}
\end{table}
\begin{figure}[htpb]
\begin{center}
\includegraphics[width=0.6\textwidth]{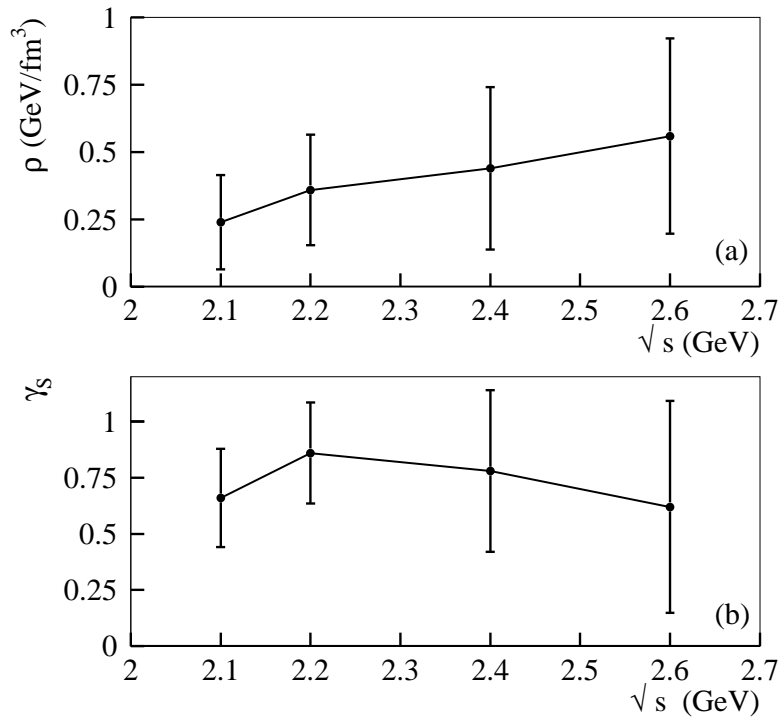}
\caption{\small{Upper panel: best fit energy density $\rho$ as a function of centre-of-mass
energy. Lower panel: best fit $\gamma_{S}$ values. Lines are drawn to guide the eye.}}
\label{rhogamma}
\end{center}
\end{figure}

Much data in this energy interval has been lately provided by the BABAR experiment
which has measured the cross-sections of several multi-hadronic channels in \ee 
collisions at several centre-of-mass energies with the method of initial state 
radiation. We have chosen four energy points, that is $\sqrt s =$ 2.1, 2.2, 2.4 and 
2.6 GeV and added to the available BABAR measurements older measurements performed 
by experiments at \ee colliders run at the same centre-of-mass energies and collected
in a nice review paper \cite{Whalley:2003qr}. In fact, most centre-of-mass energies were 
near, but not exactly, those values. Therefore, for each chosen energy, we have 
interpolated cross sections by making a simple mean of cross sections and errors 
measured at adjacent energy values. For each 
energy point, the thereby obtained cross-section values and errors from different 
experiments have been further averaged according to the weigthed average method 
used by the Particle Data Group \cite{pdg2010}, including error rescaling by 
$\chi^2/dof$ in case of large discrepancy. The full set of channels can be read 
through in tables~\ref{2.1table}, \ref{2.2table}, \ref{2.4table}, \ref{2.6table} in 
Appendix C.

In order to compare the calculation with the data of exclusive channels rate, given 
in terms of a cross section, we have introduced a normalization free parameter 
$A(\sqrt{s})$:
\begin{equation}\label{epem1}
\sigma_{\Nj} = A(\sqrt{s})\omega_{\Nj} 
\end{equation}
Finally, we have fitted all available measurements of exclusive channels rates 
at a given energy to the SHM with four free parameters: $\rho$, $\gamma_S$, $A$ and 
$I_0$. The fit minimizes the $\chi^2$:
\begin{equation}\label{chi2}
 \chi^2 = \sum_{\Nj_{\rm measured}} \frac{\left( \sigma_{\Nj}^{\rm exp} - 
 \sigma_{\Nj}^{\rm theo}\right)^2}{\Delta^2_{\rm exp}+ \Delta^2_{\rm theo}} \; .
\end{equation}
where the sum runs over measured channels; $\Delta_{\rm exp}$ is the experimental
error and $\Delta_{\rm theo}$ is the theoretical uncertainty on the cross sections 
respectively. The latter is the sum in quadrature of the statistical error, owing 
to the finite statistics in Monte-Carlo integration and the systematic error stemming
from the uncertainty on branching ratios of resonances. This error has been estimated 
at each step of the $\chi^2$ minimization by varying the branching ratios by their 
errors quoted in the Particle Data Book (or by making an educated guess if missing) 
and calculating the difference between the theoretical value of the probability
(\ref{total}) before and after the variation. This additional uncertainty generally
overcomes the experimental error for ligth particle channels with $>2$ particles, as 
the number of contributing channels with resonances is large (as it can be seen in 
tables~\ref{2.1table}, \ref{2.2table}, \ref{2.4table} and \ref{2.6table} in Appendix
C).

\begin{figure}[htpb]
\begin{center}
\includegraphics[width=0.6\textwidth]{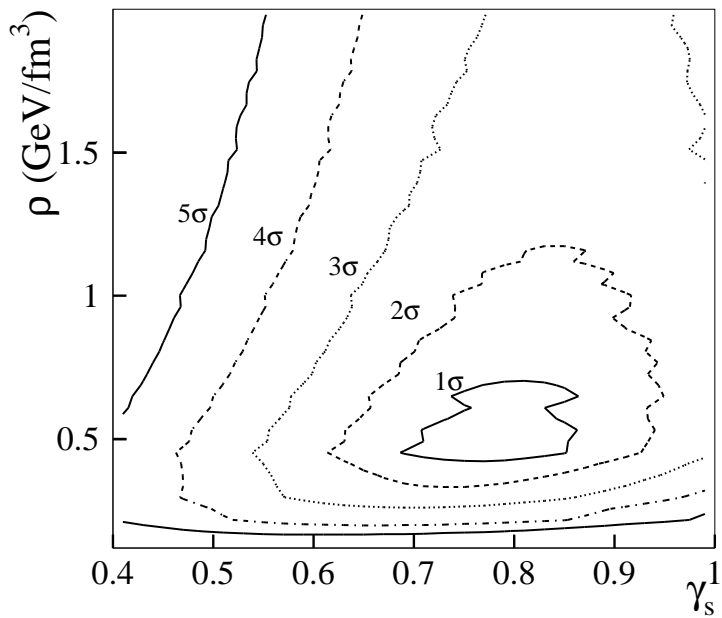}
\caption{\small{Contour plot of the $\chi^2$ for $\sqrt s = 2.4\;$ GeV.}}
\label{contplot}
\end{center}
\end{figure}

The fit procedure is as follows: for each energy point a two-dimensional grid in 
the parameters $\rho$ and $\gamma_S$ is set, with 50 divisions in each direction
and range [0.04-2] GeV/fm$^3$, [0.02-1] respectively. At each point of the grid, a 
minimization of the $\chi^2$ is carried out to determine the parameters $I_0$
and $A$. The point grid where the minimum among all minima lies has been taken as
the best fit. The error on the parameters $\rho$ and $\gamma_S$ has been estimated
graphically from the contour $\chi^2 = \chi^2_{\rm min}+1$ (see fig.~\ref{contplot}).

\begin{figure}[htpb]
\begin{center}
\includegraphics[width=0.6\textwidth]{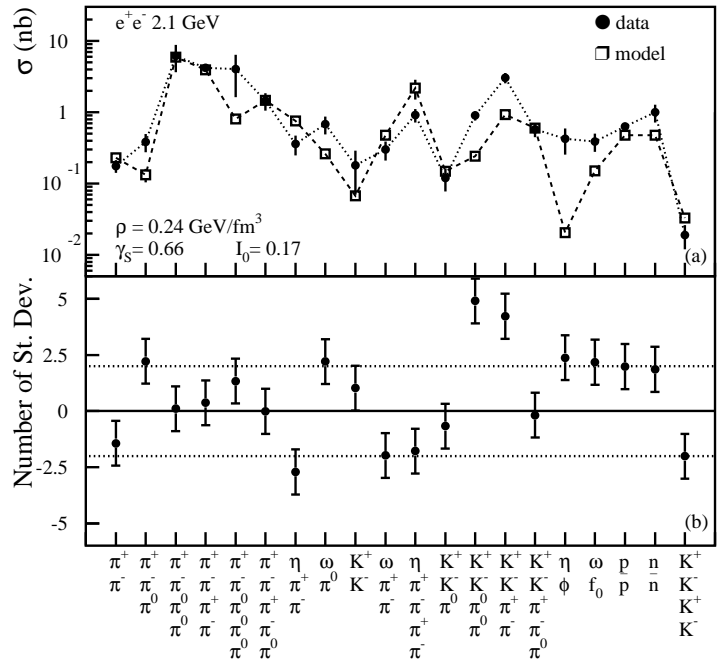}
\caption{\small{Upper panel: exclusive cross sections of various channel in \ee 
collisions at 2.1 GeV. The data points, shown as full dots, are weighted averages 
of available measurements (see text) while the fitted points are shown as open 
squares. The lines have been drawn to guide the eye. Lower panel: fit residual 
distribution.}}
\label{2.1best}
\end{center}
\end{figure}

\begin{figure}[htpb]
\begin{center}
\includegraphics[width=0.6\textwidth]{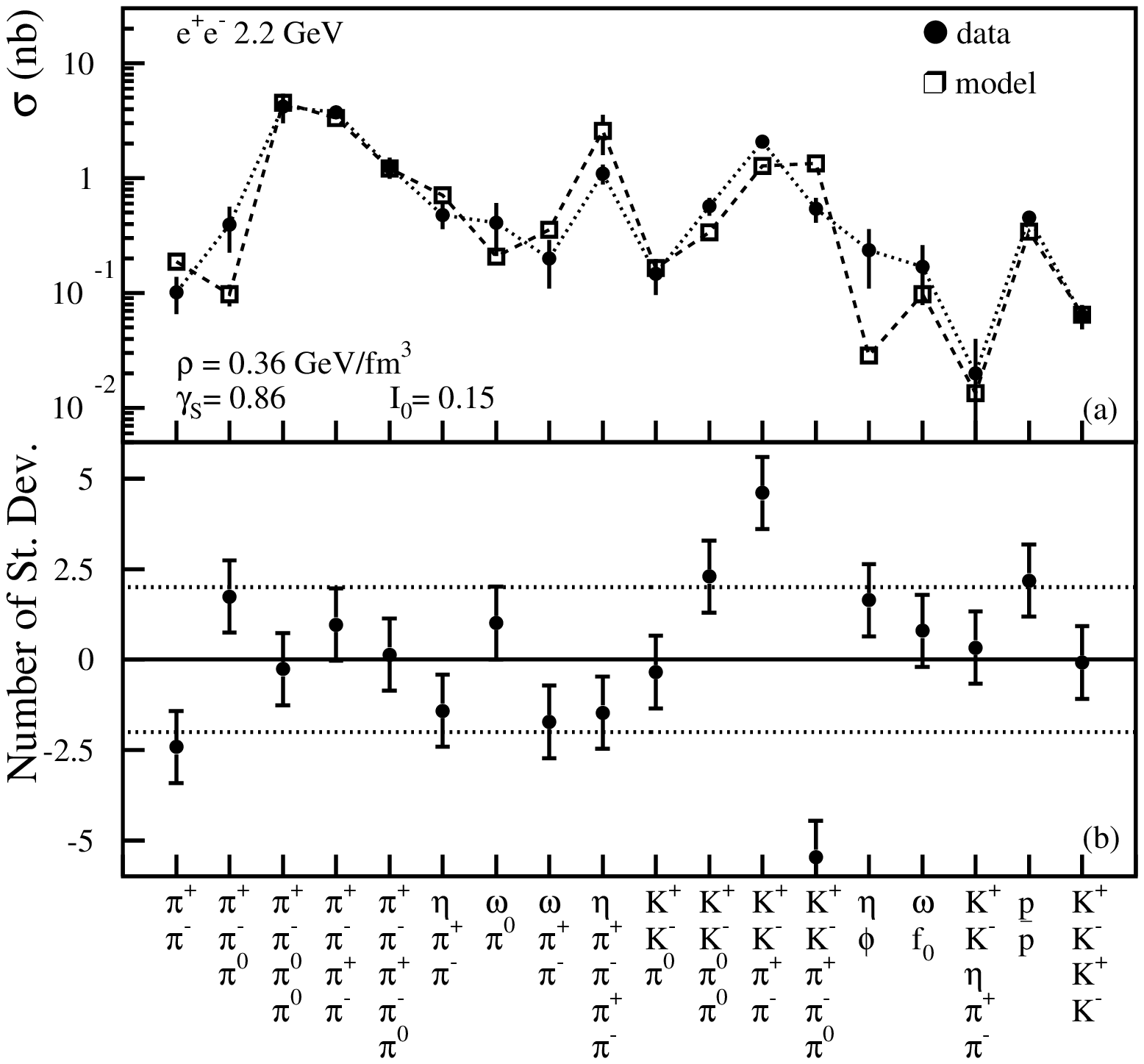}
\caption{\small{Upper panel: exclusive cross sections of various channel in \ee 
collisions at 2.2 GeV. The data points, shown as full dots, are weighted averages 
of available measurements (see text) while the fitted points are shown as open 
squares. The lines have been drawn to guide the eye. Lower panel: fit residual 
distribution.}}
\label{2.2best}
\end{center}
\end{figure}

The fit results are shown in summarized in table~\ref{fitsumm} and in fig.~\ref{rhogamma}.
The comparison of the fitted rates with the data is shown in figs.~\ref{2.1best}, 
\ref{2.2best}, \ref{2.4best}, \ref{2.6best} and in tables~\ref{2.1table}, \ref{2.2table}, 
\ref{2.4table}, \ref{2.6table} in Appendix C for $\sqrt{s}=$ 2.1, 2.2, 2.4, 2.6 GeV 
respectively. 

\begin{figure}[htpb]
\centering
\subfigure[]
{\includegraphics[scale=0.6]{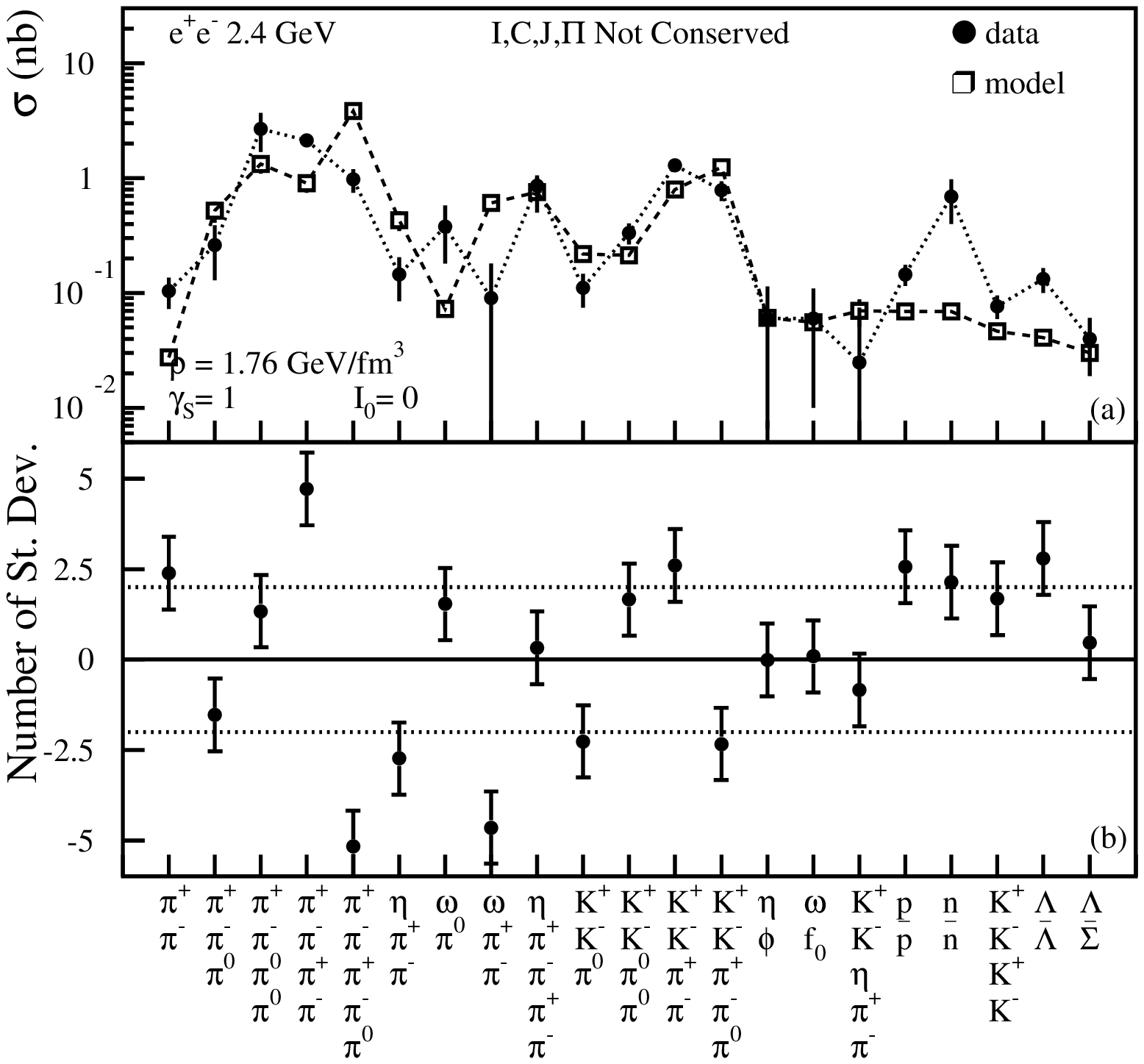}
\label{2.4first}}
\subfigure[]
{\includegraphics[scale=0.6]{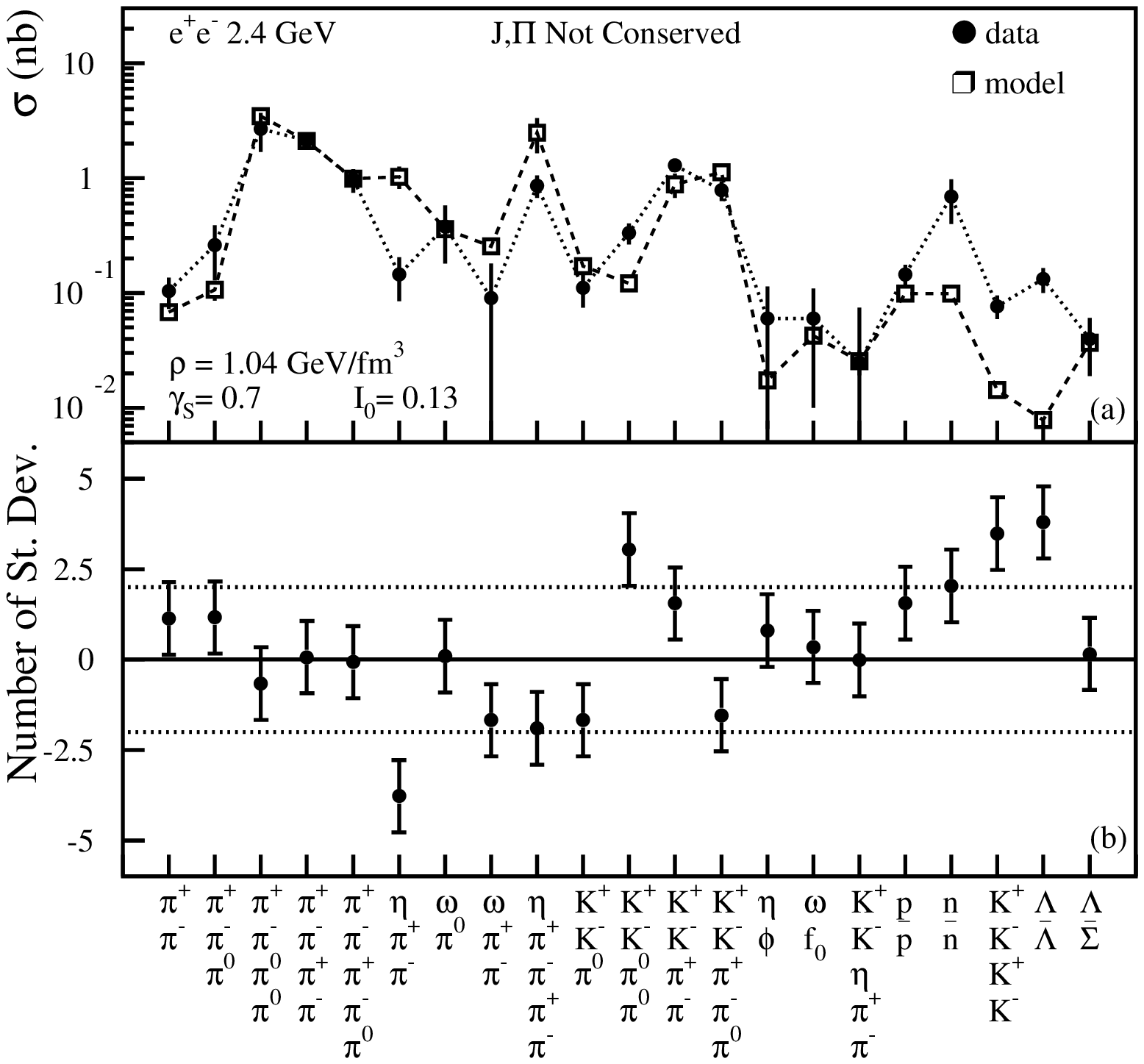}
\label{2.4second}}
\caption[]{Fits to exclusive channels in \ee collisions at 2.4 with conservation laws
turned off. The data points, shown as full dots, are weighted averages 
of available measurements (see text) while the fitted points are shown as open 
squares. Panel (a): fit with only energy-momentum conservation. Panel (b): fit without 
angular momentum conservation.}
\end{figure}

\subsection{Test of conservation laws}
\label{testlaws}

To appraise the role of the various constraints and the features of the statistical 
model, it is worth making the fit by switching off the conservation laws in turn. We 
have therefore minimized the $\chi^2$ with the same aforementioned procedure, at a
single energy point $\sqrt s =$ 2.4 GeV, in two different modes. 

\begin{figure}[htpb]
\begin{center}
\includegraphics[width=0.6\textwidth]{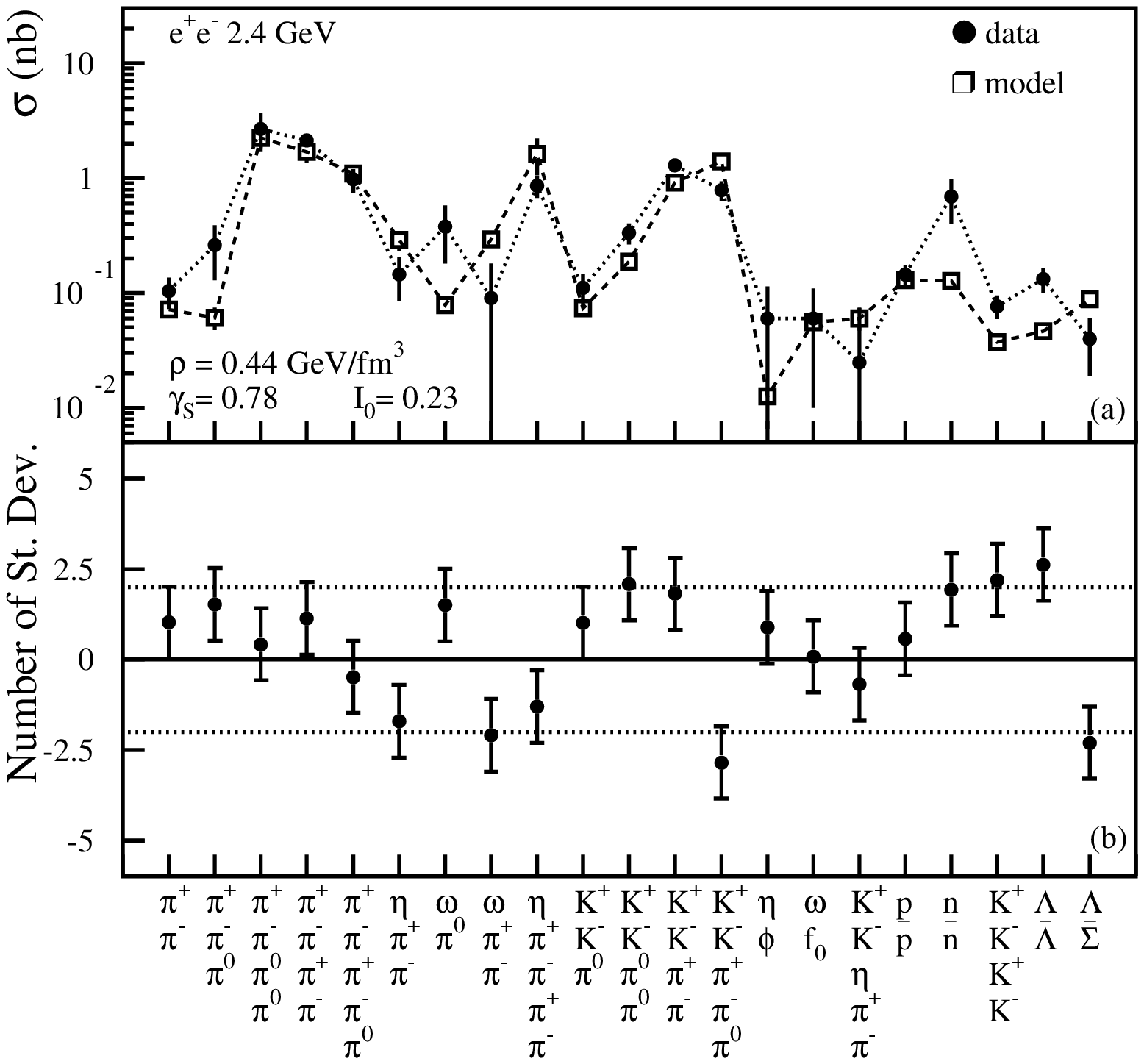}
\caption{\small{Upper panel: exclusive cross sections of various channel in \ee 
collisions at 2.4 GeV. The data points, shown as full dots, are weighted averages 
of available measurements (see text) while the fitted points are shown as open 
squares. The lines have been drawn to guide the eye. Lower panel: fit residual 
distribution.}}
\label{2.4best}
\end{center}
\end{figure}

In the first mode we have kept only the conservation of energy and momentum, disregarding 
angular momentum, parity and internal symmetries. The result is shown in fig.~\ref{2.4first},
where a consistent worsening of the fit quality can be seen ($\chi^2$/dof = 133/17), 
especially looking at the residual distribution. The fit quality significantly improves 
by turning on the internal symmetries ($\chi^2$/dof = 74.5/17). Finally, restoring the
angular momentum and parity conservation, one obtains the best fit shown in 
fig.~\ref{2.4best} ($\chi^2$/dof = 55.4/17). This further improvement indicates that
angular momentum conservation plays an important role and this was indeed expected as
this is a very effective mechanism in modulating the rate of a channel for a spacially 
extended source (the well known centrifugal barrier effect), which is one of 
the crucial assumption of the statistical model.

\begin{figure}[htpb]
\begin{center}
\includegraphics[width=0.6\textwidth]{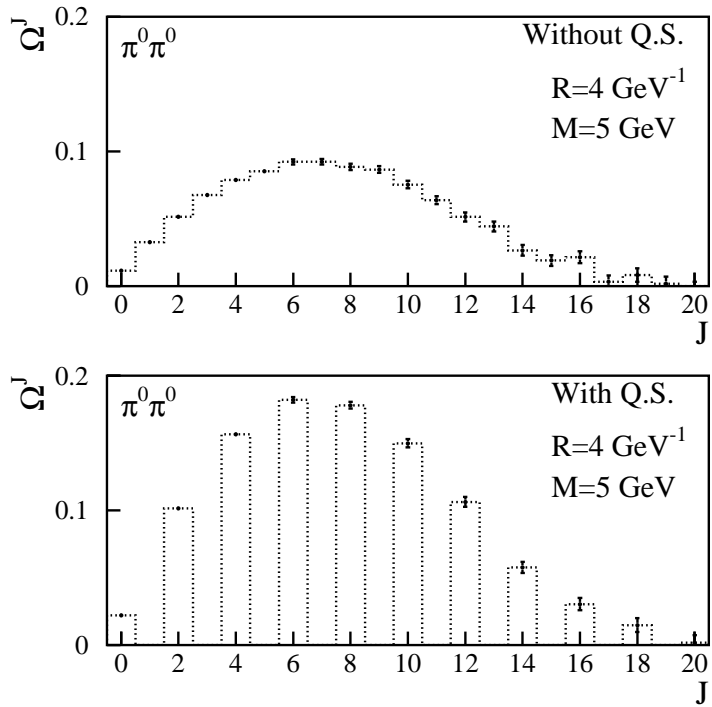}
\caption{\small{Normalized microcanonical weights of the channel $\pi^0 \pi^0$ as 
a function of the cluster spin $J$ in the Boltzmann statistics (upper panel) and 
quantum statistics (lower panel) case. The cluster has been taken spherical
in shape with radius $R=4$~GeV$^{-1}$ and a mass $M=5$~GeV. Isospin, parity, 
$C$-parity conservation are turned off.}}
\label{spectra}
\end{center}
\end{figure}

To highlight this effect, we have calculated the microcanonical channel weight for 
the channel $\pi^0 \pi^0$ for different spin $J$ of the cluster. Assuming a spherical
shape and switching off parity and internal symmetries (isospin and charge conjugation)
we have obtained the normalized microcanonical channel weights shown in fig.~(\ref{spectra}), 
for a cluster mass $M$ of $5$ GeV and a radius of $R=4$~GeV$^{-1}$. As expected,
the maximal value of the microcanonical channel weight is located around the angular 
momentum $J \sim {\rm p} R \sim 10 $, $\rm p$ being the momentum of the pion.  
This calculation also served as a consistency check for our numerical code as the
sum of all microcanonical channel weights for all $J$ was found to reproduce the
simple two-body phase space with only energy-momentum conservation.

\begin{figure}[htpb]
\begin{center}
\includegraphics[width=0.6\textwidth]{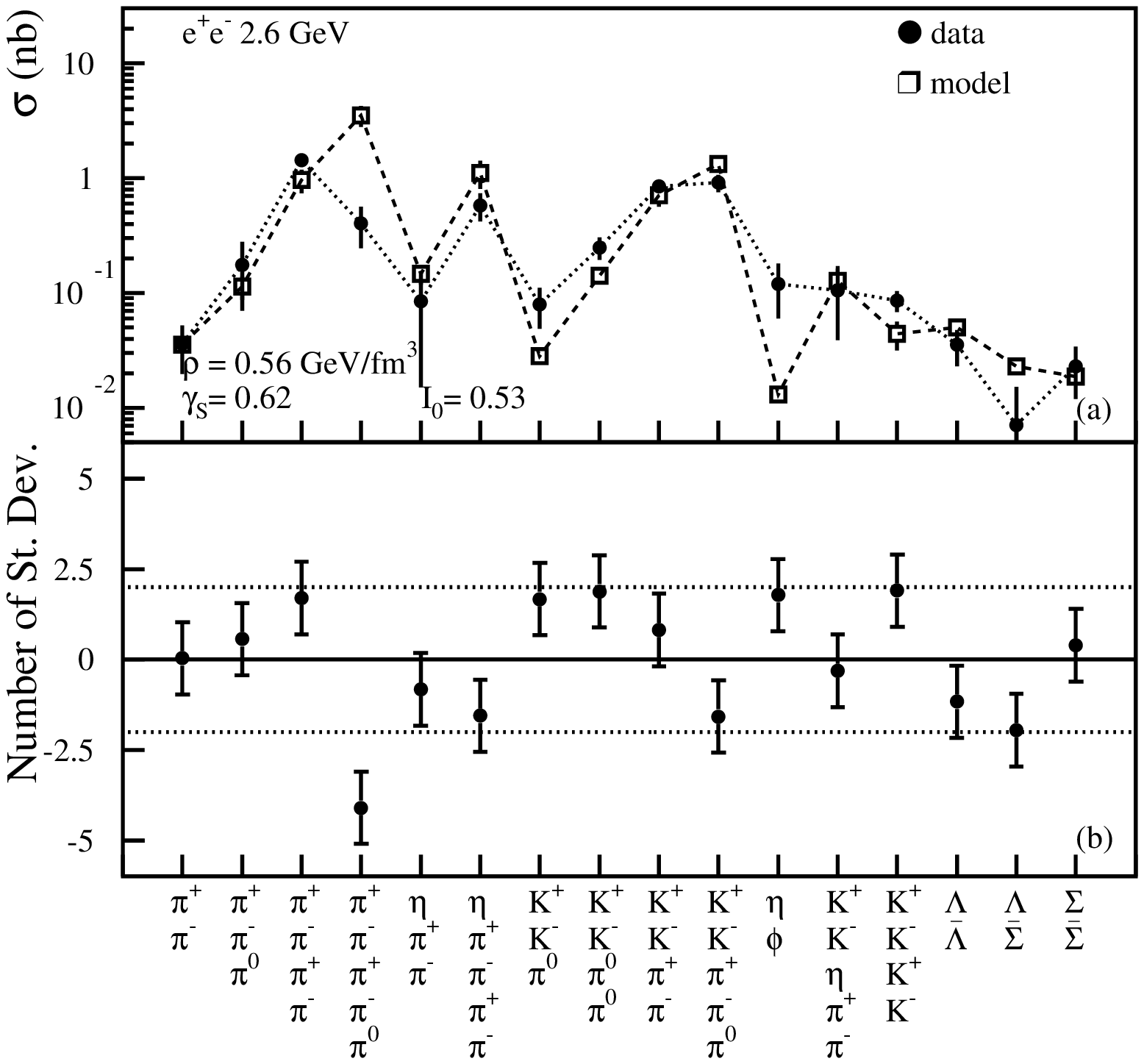}
\caption{\small{Upper panel: exclusive cross sections of various channel in \ee 
collisions at 2.6 GeV. The data points, shown as full dots, are weighted averages 
of available measurements (see text) while the fitted points are shown as open 
squares. The lines have been drawn to guide the eye. Lower panel: fit residual 
distribution.}}
\label{2.6best}
\end{center}
\end{figure}

\section{Discussion and conclusions}\label{concl}

Although the fit quality is not perfect in terms of statistical test (see table
\ref{fitsumm}), we can fairly conclude, looking at figs.~\ref{2.1best}, \ref{2.2best},
\ref{2.4best}, \ref{2.6best} that the statistical hadronization model is able to 
satisfactorilty reproduce most exclusive multi-hadronic channels measured in \ee 
collisions at low energy. Especially at 2.4 GeV, all measured channel rates lie 
within 2.5 standard deviations from the model values, which is quite remarkable 
taking into account the obvious fact that exclusive channels are a very stringent 
test for any model, certainly much more than inclusive multiplicities, and that
the fits were done with only 4 free parameters.

To fairly judge the quality of the agreement between model and data, it is worth
keeping in mind that the analysis we have presented in this work still relies on
several approximations, so that one may hope that a more thorough calculation 
will result in a better agreement between model and data. The main approximation 
introduced in this calculation is the assumption of validity of the hadron-resonance 
gas model for exclusive channels and at finite volume, what is in fact granted 
only for fully inclusive quantities and in the thermodynamic limit. Moreover, 
the assumed symmetry SU(2)$_{\rm iso}\otimes$U(1)$_{\rm strange}$ may not be a 
sufficiently realistic scheme. 

The statistical model nicely matches one of the main features of the data, namely 
the gross dependence of the channel rate on the number of particles $N$; this 
kind of hyerarchy can be clearly observed in all fits (see e.g. fig.~\ref{2.4best}) 
and the SHM is able to reproduce this behaviour because of the approximate dependence 
of the rate on $V^N$, as in formula (\ref{microchapp}). Furthermore, the more 
conservation laws are included, the more the fit improves and the more the model 
predictions approach the actual rates. This is a certainly a good point for the 
SHM, especially as far as angular momentum is concerned, as has been discussed in 
Subsect.~\ref{testlaws}.

Overall, the most interesting outcome of the analysis are the values of the fitted
energy density $\rho$ and strangeness suppression parameter $\gamma_S$, shown in
fig.~\ref{rhogamma}, around 0.5 GeV/fm$^3$ and 0.7 respectively. These values are 
essentially the same obtained with the analysis of inclusive hadronic multiplicities
at high energy \cite{review} \footnote{For a neutral hadron-resonance gas in the
canonical ensemble, the energy density of 0.4 GeV/fm$^3$ approximately corresponds
to a temperature of 160 MeV} and this confirms the consistency of the statistical 
approach to hadronization. While the origin of extra-strangeness suppression is 
not clear (an interesting proposal was put forward in ref.~\cite{bcms}) the idea 
of hadronization as a process occurring at a critical energy density which uniformly 
populates the available phase space is certainly reinforced by the observation that 
this seems to happen at universal values of the parameters, at high as well as at 
low energy.   

The interpretation of the statistical equilibrium in hadronization is an open issue
and there are several proposals. One of the authors (F.B.) favours the idea of
a quantum-chaotic effect (known as Berry's conjecture) owing the the strong non-
linearity of QCD in the non-perturbative regime. 


\section*{Appendix A}

The charge-conjugation operator ${\sf C}$, when acting on a charged light-flavoured 
meson belonging to an isotriplet may generate an arbitrary phase. Fixing this 
phase is essential for our definition of the projector (\ref{eqn:isoscp1}). For this 
purpose, let us define the operator $\widehat{{\sf G}}$ as:
\begin{equation}\label{app4a_1} 
\widehat{{\sf G}}=\widehat{{\sf C}}\; \e^{\i \pi \widehat{I}_2} 
\end{equation}
which is known as {\em G-parity operator}. When applied to a state with third component 
of isospin $I_3$, this amounts to first flipping $I_3 \rightarrow -I_3$ and reversing
the process, $-I_3 \rightarrow I_3$. Therefore, an eigenvector of $\widehat{\sf I}_3$ 
is also an eigenvector of the G-parity operator. For an isotriplet like pions':
\begin{eqnarray}\label{app4a_2} 
&&\widehat{{\sf G}}\ket{\pi^\pm}=-\widehat{\sf C}\ket{\pi^\pm} \\ \nonumber
&&\widehat{{\sf G}}\ket{\pi^0}=\pm\ket{\pi^0} \; . 
\end{eqnarray}
because $\e^{i \pi \widehat{I}_2}$ results in a factor $-1$ when applied to
any pion state. Now, since $\ket{\pi^0}$ is an eigenstate of $\widehat{\sf C}$ 
with eigenvalue $+1$, we define:
\begin{equation}
\label{app4a_2x0x} 
\widehat{{\sf C}}\ket{\pi^\pm}=-\ket{\pi^\mp} \; .
\end{equation}
so as to G-parity to yield the same eigenvalue for all members of the isotriplet.

\section*{Appendix B}

One of the important steps in the numerical evaluation of the microcanonical
channel weight is the computation of: 
\begin{equation}
\label{app4_0}
\mathcal{I}_{\rhov}^{\Nj}\left( I, I_3\right) \equiv \left[ \prod_{j=1}^{K} 
\bra{ I_j, \{ I_3^{n_j}\}} \right] \ket{ I,I_3 } \bra{ I,I_3 } 
\left[ \prod_{j=1}^{K} \ket{ \{I_j, I_3^{\rho_{j}(n_j)}\} } \right] 
\end{equation}
We describe here a method to compute the more general expression:
\begin{equation}
\label{app4_1} 
\left\{ \bra{I^1,I^{1}_3} \bra{I^2,I_3^{2}}\cdots \bra{I^N,I_3^{N}} \right\} 
\ket{ I,I_3 }\bra{ I,I_3 } \left\{ \ket{I^1,I_3^{1}\,'} \ket{I^2,I_3^{2}\,'} 
\cdots \ket{I^N,I_3^{N}\,'} \right\} \qquad
\end{equation}
where
\begin{equation}
\label{app4_2} 
\ket{I^1,I_3^{1}} \ket{I^2,I_3^{2}} \cdots \ket{I^N,I_3^{N}}
\end{equation}
is a generic multi-particle isospin state. A closed analytical formula for (\ref{app4_0}),
as a finite sum, has been found in ref.~\cite{ceriso4}. We have not used that formula, 
yet, in the following, we will closely follow the notations therein.

In order to calculate the projection of the state (\ref{app4_2}) onto the subspace
with total isospin $\ket{II_3}$, one should use a system of base-vectors in isospin 
space where $I$ is diagonal. The choice of such a basis is equivalent to the choice 
of a coupling scheme for the $I_i$~\cite{ceriso4}; we can choose, for instance, a 
base where:
\begin{eqnarray}
\label{app4_3} 
&&(I^{12})^2=(I^1+I^2)^2 \\ \nonumber
&&(I^{123})^2=(I^1+I^2+I^3)^2 \\ \nonumber
&&\ldots
\end{eqnarray}
are diagonal.

A base vector for this scheme is denoted by:
\begin{equation}
\label{app4_4} 
\ket{I^{12},I^{123},\ldots, I,I_3}
\end{equation}
where $I^{1}$, $I^{2}$, etc., are known. Both sets (\ref{app4_4}) and (\ref{app4_2}) 
are complete and they are connected by a unitary transformation:
\begin{eqnarray}
\label{app4_5} 
 &&\ket{I^{12},I^{123},\ldots, I,I_3} \\ \nonumber \\ \nonumber
 &&=\sum_{I_3^1, \ldots, I_3^N} \left\{ \bra{I^1,I_3^{1}}\cdots \bra{I^N,I_3^{N}} 
 \right\} \ket{I^{12},I^{123},\ldots, I,I_3} \left\{ \ket{I^1,I_3^{1}}  \cdots 
 \ket{I^N,I_3^{N}}\right\}
\end{eqnarray}
The coefficient $\left\{ \bra{I^1,I_3^{1}}\cdots \bra{I^N,I_3^{N}} 
\right\} \ket{I^{12},I^{123},\ldots, I,I_3}$ is called a {\em recoupling coefficient} 
and is a product of Clebsch-Gordan coefficients:
\begin{equation}
\label{app4_6} 
\left\{ \bra{I^1,I_3^{1}}\cdots \bra{I^N,I_3^{N}} 
\right\} \ket{I^{12},I^{123},\ldots, I,I_3} =C^{I^1 I^2 I^{12}}_{I_3^{1}I_3^{2}
 (I_3^{1}+I_3^{2})} \cdots 
C^{I^{12\ldots N-1}I^{N} I}_{(I_3^{1}+I_3^{2}+\ldots I_3^{N-1} ) I_3^{N} I_3} \qquad
\end{equation}
By using eq.~(\ref{app4_6}) one can rewrite (\ref{app4_1}) as:
\begin{eqnarray}
\label{app4_7} 
&&\left\{ \bra{I^1,I^{1}_3} \cdots \bra{I^N,I_3^{N}} \right\} \ket{ I,I_3} 
\bra{ I,I_3 } \left\{ \ket{I^1,I_3^{1}\,'} \cdots \ket{I^N,I_3^{N}\,'} \right\}  
 \\ \nonumber \\ \nonumber
&&\qquad = \sum_{I^{12}I^{123}\ldots I^{12\ldots N-1}} \left\{ \bra{I^1,I_3^{1}}
 \cdots \bra{I^N,I_3^{N}} 
 \right\}   \ket{I^{12},I^{123},\ldots, I,I_3} \\ \nonumber \\ \nonumber
 &&\qquad  \times\bra {I^{12},I^{123},\ldots, I,I_3}
\left\{ \ket{I^1,I_3^{1}\,'}  \cdots \ket{I^N,I_3^{N}\,'}\right\}  \\ \nonumber 
 \\ \nonumber
&&\qquad   = \sum_{I^{12}I^{123}\ldots I^{12\ldots N-1}} 
 C^{I^1 I^2 I^{12}}_{I_3^{1}I_3^{2}(I_3^{1}+I_3^{2})} \cdots 
C^{I^{12\ldots N-1}I^{N} I}_{(I_3^{1}+I_3^{2}+\ldots I_3^{N-1} ) I_3^{N} I_3}  
 \\ \nonumber \\ \nonumber
&&\qquad   \times C^{I^1 I^2 I^{12}}_{I_3^{1}\,' I_3^{2}\,'(I_3^{1}\,'+I_3^{2}\,')} \cdots 
C^{I^{12\ldots N-1}I^{N} I}_{(I_3^{1}\,'+I_3^{2}\,'+\ldots I_3^{N-1}\,' ) I_3^{N}\,' I_3}
\end{eqnarray}
of course, if one of the Clebsch-Gordan coefficient is vanishing, the corresponding 
term in the previous sum is vanishing too.  

\begin{figure}[htpb]
\begin{center}
\includegraphics[width=1.0\textwidth]{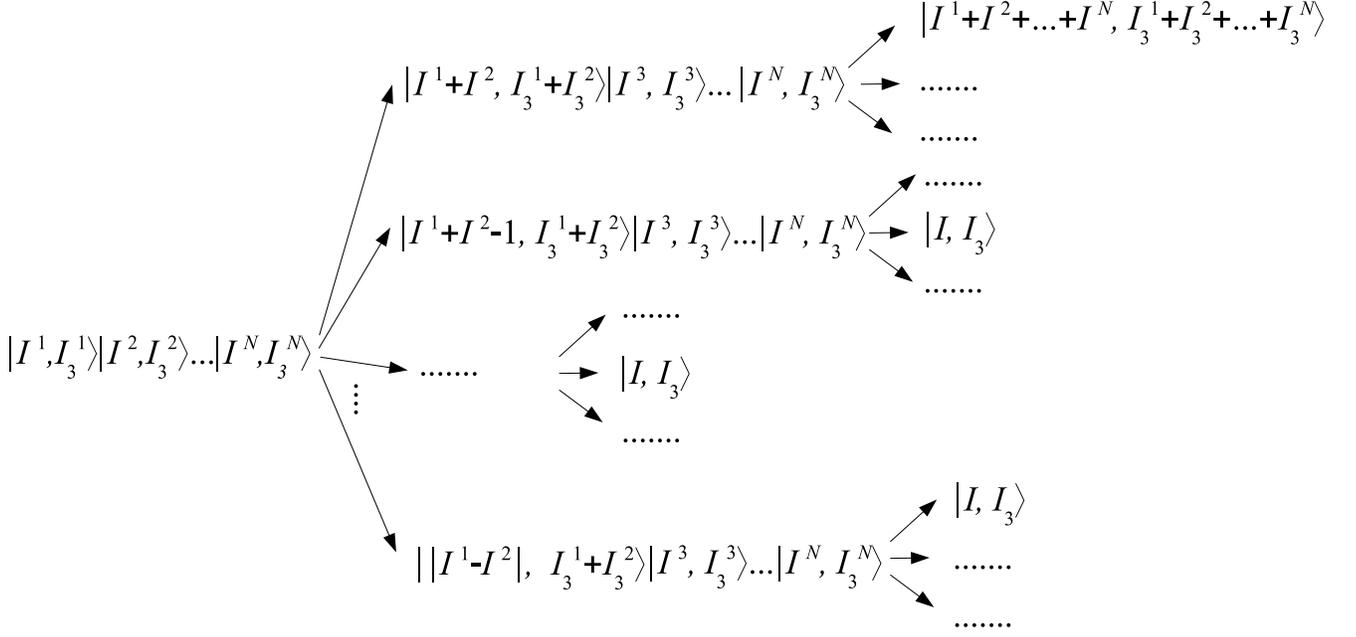}
\caption{\small{Tree diagram for the calculation of isospin matrix elements.}}
\label{isospintree}
\end{center}
\end{figure}

We have implemented a recursive numerical method to calculate the above expression.
By iterating the coupling scheme in (\ref{app4_3}) one can build a tree 
diagram, shown in fig.~(\ref{isospintree}) from left to right, where for each recoupling 
step, the highest value of the resulting isospin is put on top and the other values are 
sorted in decreasing order.

To each branch of the tree a numerical coefficient is associated, which is the 
recursive product of the Clebsch-Gordan coefficient corresponding to the decomposition
which generated that branch and the numerical coefficient of the generating branch.  
At the righmost end of the tree, one finds all possible values of the global 
isospin which can be obtained by coupling the set $I^1, I_3^1; \ldots;I^N,I_3^N$ 
and the corresponding recoupling coefficients $\braket{I,I_3}{I^1,I_3^{1},I^2,I_3^{2}, 
\ldots I^N,I_3^{N}}$. 

The main advantage of this algorithm resides in the possibility of a simple 
recursive implementation. Moreover, the CPU time needed for the tree exploration
can be easily reduced by simply switching off the recursion for branches with 
vanishing Clebsch-Gordan coefficients.

\section*{Appendix C}

In this Appendix we have collected the tables with best fit values and the experimental
values of the cross sections of exclusive channels. 

\begin{table}[htpb]
\label{2.1table}
\begin{center}
\begin{tabular}{|l|c|c|c|}
\hline
\multicolumn{4}{|c|}{$\sqrt{s}=2.1$~GeV} \\ 
\hline
channel & $\sigma_{SHM}$~(nb) & $\sigma_{exp.}$~(nb) & References \\ 
\hline
$ \pi^+ \pi^- $ & $ 0.229 \pm 0.003 \pm 0 $ & $ 0.177 \pm 0.036 $ & \cite{:2009fg,Whalley:2003qr} \\   
\hline
$ \pi^+ \pi^- \pi^0 $ & $ 0.133 \pm 0.002 \pm 0.03 $ & $ 0.385 \pm 0.11 $ & \cite{Whalley:2003qr,Aubert:2004kj} \\   
\hline
$ \pi^+ \pi^- \pi^0 \pi^0 $ & $ 5.93 \pm 0.06 \pm 0.5 $ & $ 6.2 \pm 2.57 $ & \cite{Whalley:2003qr} \\   
\hline
$ \pi^+ \pi^- \pi^+ \pi^- $ & $ 3.96 \pm 0.02 \pm 0.4 $ & $ 4.18 \pm 0.45^{*} $ & \cite{Whalley:2003qr,Aubert:2005eg} \\   
\hline
$ \pi^+ \pi^- \pi^0 \pi^0 \pi^0 $ & $ 0.809 \pm 0.02 \pm 0.05 $ & $ 4.03 \pm 2.41 $ & \cite{Whalley:2003qr} \\   
\hline
$ \pi^+ \pi^- \pi^+ \pi^- \pi^0 $ & $ 1.46 \pm 0.01 \pm 0.07 $ & $ 1.46 \pm 0.4^{*} $ & \cite{Whalley:2003qr,Aubert:2007ef} \\   
\hline
$ \eta \pi^+ \pi^- $ & $ 0.758 \pm 0.02 \pm 0.09 $ & $ 0.36 \pm 0.11 $ & \cite{Whalley:2003qr,Aubert:2007ef} \\   
\hline
$ \omega \pi^0 $ & $ 0.261 \pm 0.004 \pm 0 $ & $ 0.68 \pm 0.19 $ & \cite{Whalley:2003qr} \\   
\hline
$ K^+ K^- $ & $ 0.0673 \pm 0.0007 \pm 0 $ & $ 0.18 \pm 0.11 $ & \cite{Whalley:2003qr} \\   
\hline
$ \omega \pi^+ \pi^- $ & $ 0.478 \pm 0.002 \pm 0.001 $ & $ 0.3 \pm 0.09 $ & \cite{Whalley:2003qr,Aubert:2007ef} \\   
\hline
$ \eta \pi^+ \pi^- \pi^+ \pi^- $ & $ 2.18 \pm 0.02 \pm 0.7 $ & $ 0.915 \pm 0.2 $ & \cite{Aubert:2007ef} \\   
\hline
$ K^+ K^- \pi^0 $ & $ 0.148 \pm 0.005 \pm 0.0009 $ & $ 0.12 \pm 0.042 $ & \cite{Whalley:2003qr,Aubert:2007ur} \\   
\hline
$ K^+ K^- \pi^0 \pi^0 $ & $ 0.242 \pm 0.002 \pm 0.006 $ & $ 0.9 \pm 0.134 $ & \cite{Aubert:2007ur} \\   
\hline
$ K^+ K^- \pi^+ \pi^- $ & $ 0.929 \pm 0.003 \pm 0.05 $ & $ 3.05 \pm 0.5^{*} $ & \cite{Whalley:2003qr,Aubert:2007ur} \\   
\hline
$ K^+ K^- \pi^+ \pi^- \pi^0 $ & $ 0.6 \pm 0.006 \pm 0.03 $ & $ 0.575 \pm 0.13 $ & \cite{Aubert:2007ef} \\   
\hline
$ \eta \phi $ & $ 0.0206 \pm 0.0002 \pm 0 $ & $ 0.425 \pm 0.17 $ & \cite{Aubert:2007ym} \\   
\hline
$ \omega f_0 $ & $ 0.151 \pm 0.0007 \pm 0 $ & $ 0.39 \pm 0.11 $ & \cite{Aubert:2007ef} \\   
\hline
$ p \overline{p} $ & $ 0.479 \pm 0.002 \pm 0 $ & $ 0.63 \pm 0.076 $ & \cite{Whalley:2003qr} \\   
\hline
$ n \overline{n} $ & $ 0.48 \pm 0.002 \pm 0 $ & $ 1 \pm 0.28 $ & \cite{Whalley:2003qr} \\   
\hline
$ K^+ K^- K^+ K^- $ & $ 0.0331 \pm 6e-05 \pm 0 $ & $ 0.019 \pm 0.007 $ & \cite{Aubert:2007ur} \\   
\hline
\multicolumn{4}{|c|}{Not included in the fit} \\ 
\hline
$ \eta' \pi^+\pi^- $ & $0.371  \pm 0.005 \pm  0.008$ & $ 0.17 \pm 0.07^{**} $ & \cite{Aubert:2007ef} \\   
\hline   
$ \phi \pi^+\pi^- $ & $0.127  \pm 0.0009 \pm  0$ & $ 0.395 \pm 0.065^{**} $ & \cite{Aubert:2007ur} \\   
\hline 
$ K^*_0(892)K^-\pi^+ $ & $0.324  \pm 0.002 \pm  0.01$ & $ 1.07 \pm 0.08^{**} $ & \cite{Aubert:2007ur} \\   
\hline 
$ f_1 \pi^+\pi^- $ & $1.06  \pm 0.005 \pm  0.2$ & $ 0.79 \pm 0.24^{**} $ & \cite{Aubert:2007ef} \\   
\hline 
$ \omega K^+ K^- $ & $0.24  \pm 0.001 \pm  0$ & $ 0.49 \pm 0.09^{**} $ & \cite{Aubert:2007ef} \\   
\hline 
$ \phi f_0 $ & $0.089  \pm 0.0002 \pm  0$ & $ 0.41 \pm 0.08^{**} $ & \cite{Aubert:2007ur} \\   
\hline 
\multicolumn{4}{l}{\small{$^*$ Errors have been rescaled because of discrepancies between
different experiments.}} \\
\multicolumn{4}{l}{\small{$^{**}$ Errors are statistical only.}} \\ 
\end{tabular}
\end{center}
\caption{Comparison between fitted and experimental cross sections of exclusive multi-hadronic
channels in \ee collisions at $\sqrt s =$ 2.1 GeV. The errors on theoretical cross sections
$\sigma_{SHM}$ are the Monte-Carlo integration statistical error and the error owing to
uncertainty on branching ratios of resonances contributing to the channel, respectively.
The experimental cross section values have been obtained by averaging available measurements,
see text.}
\end{table}
\begin{table}\label{2.2table}
\begin{center}
\begin{tabular}{|l|c|c|c|}
\hline
\multicolumn{4}{|c|}{$\sqrt{s}=2.2$~GeV} \\ 
\hline
channel & $\sigma_{SHM}$~(nb) & $\sigma_{exp.}$~(nb) & References \\ 
\hline
$ \pi^+ \pi^- $ & $ 0.188 \pm 0.002 \pm 0 $ & $ 0.101 \pm 0.0359 $ & \cite{:2009fg} \\   
\hline
$ \pi^+ \pi^- \pi^0 $ & $ 0.0971 \pm 0.003 \pm 0.02 $ & $ 0.395 \pm 0.17 $ & \cite{Aubert:2004kj} \\   
\hline
$ \pi^+ \pi^- \pi^0 \pi^0 $ & $ 4.53 \pm 0.03 \pm 0.4 $ & $ 4.2 \pm 1.2 $ & \cite{Whalley:2003qr} \\   
\hline
$ \pi^+ \pi^- \pi^+ \pi^- $ & $ 3.34 \pm 0.1 \pm 0.3 $ & $ 3.73 \pm 0.2 $ & \cite{Whalley:2003qr,Aubert:2005eg} \\   
\hline
$ \pi^+ \pi^- \pi^+ \pi^- \pi^0 $ & $ 1.21 \pm 0.01 \pm 0.06 $ & $ 1.25 \pm 0.26 $ & \cite{Aubert:2007ef} \\   
\hline
$ \eta \pi^+ \pi^- $ & $ 0.712 \pm 0.007 \pm 0.1 $ & $ 0.48 \pm 0.12 $ & \cite{Aubert:2007ef} \\   
\hline
$ \omega \pi^0 $ & $ 0.207 \pm 0.003 \pm 0 $ & $ 0.41 \pm 0.2 $ & \cite{Whalley:2003qr} \\   
\hline
$ \omega \pi^+ \pi^- $ & $ 0.355 \pm 0.002 \pm 0.0008 $ & $ 0.2 \pm 0.09 $ & \cite{Whalley:2003qr,Aubert:2007ef} \\   
\hline
$ \eta \pi^+ \pi^- \pi^+ \pi^- $ & $ 2.57 \pm 0.03 \pm 1 $ & $ 1.09 \pm 0.215 $ & \cite{Aubert:2007ef} \\   
\hline
$ K^+ K^- \pi^0 $ & $ 0.166 \pm 0.02 \pm 0.001 $ & $ 0.146 \pm 0.05 $ & \cite{Aubert:2007ym} \\   
\hline
$ K^+ K^- \pi^0 \pi^0 $ & $ 0.34 \pm 0.002 \pm 0.008 $ & $ 0.57 \pm 0.1 $ & \cite{Aubert:2007ur} \\   
\hline
$ K^+ K^- \pi^+ \pi^- $ & $ 1.29 \pm 0.005 \pm 0.06 $ & $ 2.07 \pm 0.16 $ & \cite{Whalley:2003qr,Aubert:2007ur} \\   
\hline
$ K^+ K^- \pi^+ \pi^- \pi^0 $ & $ 1.34 \pm 0.008 \pm 0.07 $ & $ 0.54 \pm 0.13 $ & \cite{Aubert:2007ef} \\   
\hline
$ \eta \phi $ & $ 0.0285 \pm 0.0003 \pm 0 $ & $ 0.235 \pm 0.126 $ & \cite{Aubert:2007ym} \\   
\hline
$ \omega f_0 $ & $ 0.0976 \pm 0.0004 \pm 0 $ & $ 0.17 \pm 0.091 $ & \cite{Aubert:2007ef} \\   
\hline
$ K^+ K^- \eta \pi^+ \pi^- $ & $ 0.0134 \pm 0.0003 \pm 0 $ & $ 0.02 \pm 0.02 $ & \cite{Aubert:2007ef} \\   
\hline
$ p \overline{p} $ & $ 0.34 \pm 0.001 \pm 0 $ & $ 0.454 \pm 0.052 $ & \cite{Whalley:2003qr} \\   
\hline
$ K^+ K^- K^+ K^- $ & $ 0.0648 \pm 0.0003 \pm 0 $ & $ 0.0635 \pm 0.015 $ & \cite{Aubert:2007ur} \\   
\hline
\multicolumn{4}{|c|}{Not included in the fit} \\ 
\hline
$ \eta' \pi^+\pi^- $ & $0.341  \pm 0.005 \pm  0.007$ & $ 0.101 \pm 0.052^* $ & \cite{Aubert:2007ef} \\   
\hline   
$ \phi \pi^+\pi^- $ & $0.139  \pm 0.001 \pm  0$ & $ 0.27 \pm 0.055^* $ & \cite{Aubert:2007ur} \\   
\hline 
$ K^*_0(892)K^-\pi^+ $ & $0.508  \pm 0.003 \pm  0.02$ & $ 0.62 \pm 0.062^* $ & \cite{Aubert:2007ur} \\   
\hline 
$ f_1 \pi^+\pi^- $ & $1.02  \pm 0.005 \pm  0.17$ & $ 0.915 \pm 0.24^* $ & \cite{Aubert:2007ef} \\   
\hline 
$ \omega K^+ K^- $ & $0.3  \pm 0.002 \pm  0$ & $ 0.39 \pm 0.08^* $ & \cite{Aubert:2007ef} \\   
\hline 
$ \phi f_0 $ & $0.087  \pm 0.0003 \pm  0$ & $ 0.295 \pm 0.065^* $ & \cite{Aubert:2007ur} \\   
\hline 
\multicolumn{4}{l}{\small{$^*$ Errors are statistical only.}} \\ 
\end{tabular}
\end{center}
\caption{Comparison between fitted and experimental cross sections of exclusive multi-hadronic
channels in \ee collisions at $\sqrt s =$ 2.2 GeV. The errors on theoretical cross sections
$\sigma_{SHM}$ are the Monte-Carlo integration statistical error and the error owing to
uncertainty on branching ratios of resonances contributing to the channel, respectively.
The experimental cross section values have been obtained by averaging available measurements,
see text.}
\end{table}
\begin{table}\label{2.4table}
\begin{center}
\begin{tabular}{|l|c|c|c|}
\hline
\multicolumn{4}{|c|}{$\sqrt{s}=2.4$~GeV} \\ 
\hline
channel & $\sigma_{SHM}$~(nb) & $\sigma_{exp.}$~(nb) & References \\ 
\hline
$ \pi^+ \pi^- $ & $ 0.0716 \pm 0.0009 \pm 0 $ & $ 0.105 \pm 0.0323 $ & \cite{:2009fg} \\   
\hline
$ \pi^+ \pi^- \pi^0 $ & $ 0.0611 \pm 0.001 \pm 0.01 $ & $ 0.26 \pm 0.13 $ & \cite{Aubert:2004kj} \\   
\hline
$ \pi^+ \pi^- \pi^0 \pi^0 $ & $ 2.25 \pm 0.03 \pm 0.4 $ & $ 2.7 \pm 1 $ & \cite{Whalley:2003qr} \\   
\hline
$ \pi^+ \pi^- \pi^+ \pi^- $ & $ 1.68 \pm 0.02 \pm 0.3 $ & $ 2.12 \pm 0.2 $ & \cite{Aubert:2005eg} \\   
\hline
$ \pi^+ \pi^- \pi^+ \pi^- \pi^0 $ & $ 1.1 \pm 0.009 \pm 0.1 $ & $ 0.975 \pm 0.23 $ & \cite{Aubert:2007ef} \\   
\hline
$ \eta \pi^+ \pi^- $ & $ 0.289 \pm 0.01 \pm 0.06 $ & $ 0.145 \pm 0.06 $ & \cite{Whalley:2003qr,Aubert:2007ef} \\   
\hline
$ \omega \pi^0 $ & $ 0.0785 \pm 0.001 \pm 0 $ & $ 0.38 \pm 0.2 $ & \cite{Whalley:2003qr} \\   
\hline
$ \omega \pi^+ \pi^- $ & $ 0.294 \pm 0.002 \pm 0.04 $ & $ 0.09 \pm 0.09 $ & \cite{Whalley:2003qr,Aubert:2007ef} \\   
\hline
$ \eta \pi^+ \pi^- \pi^+ \pi^- $ & $ 1.64 \pm 0.02 \pm 0.6 $ & $ 0.865 \pm 0.19 $ & \cite{Aubert:2007ef} \\   
\hline
$ K^+ K^- \pi^0 $ & $ 0.0739 \pm 0.001 \pm 0.0009 $ & $ 0.111 \pm 0.036 $ & \cite{Aubert:2007ym} \\   
\hline
$ K^+ K^- \pi^0 \pi^0 $ & $ 0.188 \pm 0.002 \pm 0.007 $ & $ 0.335 \pm 0.07 $ & \cite{Aubert:2007ur} \\   
\hline
$ K^+ K^- \pi^+ \pi^- $ & $ 0.912 \pm 0.004 \pm 0.1 $ & $ 1.3 \pm 0.167^{*} $ & \cite{Whalley:2003qr,Aubert:2007ur} \\   
\hline
$ K^+ K^- \pi^+ \pi^- \pi^0 $ & $ 1.39 \pm 0.006 \pm 0.2 $ & $ 0.785 \pm 0.15 $ & \cite{Aubert:2007ef} \\   
\hline
$ \eta \phi $ & $ 0.0127 \pm 0.0002 \pm 0 $ & $ 0.06 \pm 0.0534 $ & \cite{Aubert:2007ym} \\   
\hline
$ \omega f_0 $ & $ 0.0558 \pm 0.0003 \pm 0 $ & $ 0.06 \pm 0.05 $ & \cite{Aubert:2007ef} \\   
\hline
$ K^+ K^- \eta \pi^+ \pi^- $ & $ 0.0604 \pm 0.0007 \pm 0.01 $ & $ 0.025 \pm 0.05 $ & \cite{Aubert:2007ef} \\   
\hline
$ p \overline{p} $ & $ 0.129 \pm 0.0005 \pm 0 $ & $ 0.146 \pm 0.03 $ & \cite{Whalley:2003qr} \\   
\hline
$ n \overline{n} $ & $ 0.128 \pm 0.0005 \pm 0 $ & $ 0.69 \pm 0.29 $ & \cite{Whalley:2003qr} \\   
\hline
$ K^+ K^- K^+ K^- $ & $ 0.0374 \pm 0.0001 \pm 0.0002 $ & $ 0.077 \pm 0.018 $ & \cite{Aubert:2007ur} \\   
\hline
$ \Lambda \overline{\Lambda} $ & $ 0.0464 \pm 0.0001 \pm 0 $ & $ 0.133 \pm 0.033 $ & \cite{Aubert:2007uf} \\   
\hline
$ \Lambda \overline{\Sigma} $ & $ 0.0883 \pm 0.0003 \pm 0 $ & $ 0.04 \pm 0.021 $ & \cite{Aubert:2007uf} \\   
\hline
\multicolumn{4}{|c|}{Not included in the fit} \\ 
\hline
$ \eta' \pi^+\pi^- $ & $0.125  \pm 0.003 \pm  0.003$ & $ 0.045 \pm 0.04^{**} $ & \cite{Aubert:2007ef} \\   
\hline   
$ \phi \pi^+\pi^- $ & $0.112  \pm 0.0006 \pm  0$ & $ 0.125 \pm 0.04^{**} $ & \cite{Aubert:2007ur} \\   
\hline 
$ K^*_0(892)K^-\pi^+ $ & $0.245  \pm 0.002 \pm  0.008$ & $ 0.425 \pm 0.045^{**} $ & \cite{Aubert:2007ur} \\   
\hline 
$ f_1 \pi^+\pi^- $ & $0.462  \pm 0.002 \pm  0.07$ & $ 0.557 \pm 0.17^{**} $ & \cite{Aubert:2007ef} \\   
\hline 
$ \omega K^+ K^- $ & $0.229  \pm 0.001 \pm  0.044$ & $ 0.23 \pm 0.06^{**} $ & \cite{Aubert:2007ef} \\   
\hline 
$ \phi f_0 $ & $0.0395  \pm 0.0002 \pm  0$ & $ 0.135 \pm 0.045^{**} $ & \cite{Aubert:2007ur} \\   
\hline 
\multicolumn{4}{l}{\small{$^*$ Errors have been rescaled because of discrepancies between
different experiments.}} \\ 
\multicolumn{4}{l}{\small{$^{**}$ Errors are statistical only.}} \\ 
\end{tabular}
\end{center}
\caption{Comparison between fitted and experimental cross sections of exclusive multi-hadronic
channels in \ee collisions at $\sqrt s =$ 2.4 GeV. The errors on theoretical cross sections
$\sigma_{SHM}$ are the Monte-Carlo integration statistical error and the error owing to
uncertainty on branching ratios of resonances contributing to the channel, respectively.
The experimental cross section values have been obtained by averaging available measurements,
see text.}
\end{table}
\begin{table}\label{2.6table}
\begin{center}
\begin{tabular}{|l|c|c|c|}
\hline
\multicolumn{4}{|c|}{$\sqrt{s}=2.6$~GeV} \\ 
\hline
channel & $\sigma_{SHM}$~(nb) & $\sigma_{exp.}$~(nb) & References \\ 
\hline
$ \pi^+ \pi^- $ & $ 0.0354 \pm 0.0005 \pm 0 $ & $ 0.0359 \pm 0.0162 $ & \cite{:2009fg} \\   
\hline
$ \pi^+ \pi^- \pi^0 $ & $ 0.114 \pm 0.002 \pm 0.03 $ & $ 0.175 \pm 0.105 $ & \cite{Aubert:2004kj} \\   
\hline
$ \pi^+ \pi^- \pi^+ \pi^- $ & $ 0.965 \pm 0.005 \pm 0.2 $ & $ 1.44 \pm 0.16 $ & \cite{Aubert:2005eg} \\   
\hline
$ \pi^+ \pi^- \pi^+ \pi^- \pi^0 $ & $ 3.52 \pm 0.05 \pm 0.7 $ & $ 0.405 \pm 0.16 $ & \cite{Aubert:2007ef} \\   
\hline
$ \eta \pi^+ \pi^- $ & $ 0.148 \pm 0.003 \pm 0.03 $ & $ 0.085 \pm 0.07 $ & \cite{Aubert:2007ef} \\   
\hline
$ \eta \pi^+ \pi^- \pi^+ \pi^- $ & $ 1.11 \pm 0.01 \pm 0.3 $ & $ 0.58 \pm 0.16 $ & \cite{Aubert:2007ef} \\   
\hline
$ K^+ K^- \pi^0 $ & $ 0.0283 \pm 0.001 \pm 0.0005 $ & $ 0.08 \pm 0.031 $ & \cite{Aubert:2007ym} \\   
\hline
$ K^+ K^- \pi^0 \pi^0 $ & $ 0.142 \pm 0.0006 \pm 0.01 $ & $ 0.25 \pm 0.056 $ & \cite{Aubert:2007ur} \\   
\hline
$ K^+ K^- \pi^+ \pi^- $ & $ 0.708 \pm 0.002 \pm 0.1 $ & $ 0.845 \pm 0.09 $ & \cite{Aubert:2007ur} \\   
\hline
$ K^+ K^- \pi^+ \pi^- \pi^0 $ & $ 1.32 \pm 0.005 \pm 0.2 $ & $ 0.915 \pm 0.16 $ & \cite{Aubert:2007ef} \\   
\hline
$ \eta \phi $ & $ 0.013 \pm 0.0002 \pm 0 $ & $ 0.12 \pm 0.06 $ & \cite{Aubert:2007ym} \\   
\hline
$ K^+ K^- \eta \pi^+ \pi^- $ & $ 0.129 \pm 0.001 \pm 0.04 $ & $ 0.105 \pm 0.066 $ & \cite{Aubert:2007ef} \\   
\hline
$ K^+ K^- K^+ K^- $ & $ 0.0442 \pm 0.0001 \pm 0.01 $ & $ 0.086 \pm 0.018 $ & \cite{Aubert:2007ur} \\   
\hline
$ \Lambda \overline{\Lambda} $ & $ 0.0501 \pm 0.0002 \pm 0 $ & $ 0.0355 \pm 0.0125 $ & \cite{Aubert:2007uf} \\   
\hline
$ \Lambda \overline{\Sigma} $ & $ 0.0231 \pm 0.0001 \pm 0 $ & $ 0.0071 \pm 0.0082 $ & \cite{Aubert:2007uf} \\   
\hline
$ \Sigma \overline{\Sigma} $ & $ 0.0186 \pm 6e-05 \pm 0 $ & $ 0.023 \pm 0.011 $ & \cite{Aubert:2007uf} \\   
\hline
\multicolumn{4}{|c|}{Not included in the fit} \\ 
\hline
$ \eta' \pi^+\pi^- $ & $0.073  \pm 0.001 \pm  0.02$ & $ 0.035 \pm 0.0325^* $ & \cite{Aubert:2007ef} \\   
\hline   
$ \phi \pi^+\pi^- $ & $0.199 \pm 0.001 \pm  0.03$ & $ 0.065 \pm 0.02^* $ & \cite{Aubert:2007ur} \\   
\hline 
$ K^*_0(892)K^-\pi^+ $ & $0.141  \pm 0.002 \pm  0.005$ & $ 0.29 \pm 0.032^* $ & \cite{Aubert:2007ur} \\   
\hline 
$ f_1 \pi^+\pi^- $ & $0.24  \pm 0.001 \pm  0.035$ & $ 0.3 \pm 0.12^* $ & \cite{Aubert:2007ef} \\   
\hline 
$ \omega K^+ K^- $ & $0.263  \pm 0.0009 \pm  0.09$ & $ 0.2 \pm 0.05^* $ & \cite{Aubert:2007ef} \\   
\hline 
$ \phi f_0 $ & $0.0439  \pm 0.0002 \pm  0$ & $ 0.065 \pm 0.03^* $ & \cite{Aubert:2007ur} \\   
\hline 
\multicolumn{4}{l}{\small{$^*$ Errors are statistical only.}} \\ 
\end{tabular}
\end{center}
\caption{Comparison between fitted and experimental cross sections of exclusive multi-hadronic
channels in \ee collisions at $\sqrt s =$ 2.6 GeV. The errors on theoretical cross sections
$\sigma_{SHM}$ are the Monte-Carlo integration statistical error and the error owing to
uncertainty on branching ratios of resonances contributing to the channel, respectively.
The experimental cross section values have been obtained by averaging available measurements,
see text.}
\end{table}

\end{document}